\documentclass{aa}  

\usepackage{graphicx}
\usepackage{float} %to use H in figure
\usepackage{txfonts}
\usepackage{makecell}
\usepackage{comment}
\usepackage{amsmath}
\usepackage{soul}
\usepackage[export]{adjustbox}

\begin{document}

   \title{Metallicity distributions of halo stars: do they trace the Galactic accretion history?}
   %\title{Metallicity distributions of halo stars: Whether they trace the Galactic accretion history}

   \author{A. Mori\inst{1}
          \and P. Di Matteo\inst{2}
          \and S. Salvadori\inst{1, 3}
          \and S. Khoperskov\inst{4}
          \and G. Pagnini\inst{2}
          \and M. Haywood\inst{2}
          }

   \institute{Dipartimento di Fisica e Astronomia, Università di Firenze, Via G. Sansone 1, 50019 Sesto Fiorentino, Firenze, Italy\\
              \email{alice.mori@unifi.it}
         \and 
            GEPI, Observatoire de Paris, PSL Research University, CNRS, Place Jules Janssen, 92195 Meudon, France
         \and
            INAF – Osservatorio Astrofisico di Arcetri, Largo Enrico Fermi 5, 50125 Firenze, Italy
         \and 
            Leibniz-Institut für Astrophysik Potsdam (AIP), 14482 Potsdam, Germany
             }

   \date{
        }

% \abstract{}{}{}{}{} 
% 5 {} token are mandatory
 
  \abstract
{The standard cosmological scenario predicts a hierarchical formation for galaxies. Many substructures have been found in the Galactic halo, usually identified as clumps in kinematic spaces, like the energy-angular momentum space ($E-L_z$), under the hypothesis that these quantities should be conserved during the interaction. If these clumps also feature different chemical abundances, such as the metallicity distribution function (MDF), these two arguments together (different kinematic and chemical properties) are often used to motivate their association with distinct and independent merger debris.} 
{The aim of this study is to explore to what extent we can couple kinematic characteristics and metallicities of stars in the Galactic halo to reconstruct the accretion history of the Milky Way (MW). In particular, we want to understand whether different clumps in the $E-L_z$ space with different MDFs should be associated with distinct merger debris.}
{We analysed dissipationless, self-consistent, high-resolution N-body simulations of a MW-type galaxy accreting a satellite with a mass ratio of 1:10, with different orbital parameters and different metallicity gradients, which were assigned a posteriori.}
{We confirm that accreted stars from a $\sim$1:10 mass ratio merger event redistribute in a wide range of $E$ and $L_z$, due to the dynamical friction process, and are thus not associated with a single region. Because satellite stars with different metallicities can be deposited in different regions of the $E-L_z$ space (on average the more metal-rich ones end up more gravitationally bound to the MW), this implies that a single accretion of $\sim$1:10 can manifest with different MDFs, in different regions of the $E-L_z$ space.}
{Groups of stars with different $E$, $L_z$, and metallicities may be interpreted as originating from different satellite galaxies, but our analysis shows that these interpretations are not physically motivated. 
In fact, as we show, the coupling of kinematic information with MDFs to reconstruct the accretion history of the MW can bias the reconstructed merger tree towards increasing the number of past accretions and decreasing the masses of the progenitor galaxies.}

   \keywords{Galaxies: interactions -- Galaxy: formation -- Galaxy: evolution -- Galaxy: kinematics and dynamics -- Galaxy: abundances -- Methods: numerical}
             %Galaxy: halo 
   \maketitle
%

%--------------------------------------------------------------------
\section{Introduction}
%--------------------------------------------------------------------

According to the standard cosmological model (i.e. Lambda Cold Dark Matter, $\Lambda$CDM), galaxies are supposed to grow hierarchically from small sub-units that gradually merge to make up the galaxies that we observe \citep[e.g.][]{whiterees1978}. The study of a galaxy's merging history is thus crucial to understanding how galaxies form and evolve. 
In this paper, we aim to find methods to reconstruct the hierarchical build-up of galaxies by focusing on our own Milky Way (MW), for several reasons.
First, the total stellar mass of the MW is $M_* = 6 \pm{1} \times 10^{10} M_{\odot}$ \citep{bland-hawthorngerhard2016}, which places it at the peak of the distribution of galaxy masses in the local Universe \citep[i.e. at redshifts $\leq 0.1$, see][]{vandokkum13, papovich2015A}. Thus, the merging history of a large fraction of galaxies can be understood through that of a MW-type galaxy. 
Second, the MW is the only massive galaxy for which we are able to resolve single stars of different masses, ages, and compositions, and to study their distances, motions, chemical abundances, and ages. Therefore, it is the benchmark for the study of galaxy formation and evolution.
Finally, we can now rely on an unprecedented amount of data for MW stars thanks to the ESA astrometric mission \textit{Gaia} and its complementary spectroscopic surveys. Indeed, the third \textit{Gaia} data release \citep{gaia16, gaiadr3} delivered parallaxes and proper motions for over a billion MW stars, along with full 6D phase-space information for 33 million stars \citep{katz23}. On the other hand, spectroscopic surveys like APOGEE \citep{majewski17}, GALAH \citep{desilva15, buder21}, \textit{Gaia}-ESO Survey \citep{gilmore12, randich13}, H3 \citep{conroy19a}, LAMOST \citep{zhao12} and soon 4MOST \citep{dejong19}, WEAVE \citep{jin23}, and MOONS \citep{cirasuolo20} are providing chemical abundances and radial velocities for several million stars in the Galaxy up to several kiloparsecs from the Sun. 
Understanding how to interpret these new %(and upcoming) 
data is key to shedding light on the properties of the building blocks that made up our Galaxy.

Galaxies grow their stellar mass both by forming new stars, which represent the in situ population, and by accreting nearby satellite galaxies, which bring their own stars that form the so-called ex situ populations \citep{tacconi2010, scoville2014}. These ex situ stellar populations provide information on the number and mass of accreted satellites, and thus on the build-up of the MW. 
Stars belonging to satellite galaxies in the process of being accreted are still observable in the Galactic halo, since these satellites are not totally disrupted yet, but rather form stellar streams that are identifiable in the plane of the sky. The most famous example is the Sagittarius dwarf spheroidal galaxy, which is the closest MW companion, located at $\approx 24$ kpc from the Sun \citep{ibata1994}, forming a stellar stream on a polar-like orbit \citep{newberg2002, majewski2003, antoja2020}.

The reconstruction of the past accretion\footnote{In the context of this paper, "accretion" is used to refer to mergers of any mass ratio.} events of our Galaxy is not trivial, since nowadays the accreted stars associated with those events are completely mixed with the ones formed in situ. The accreted stars, indeed, lose their spatial coherence and after a few Gyr they cannot be detected anymore as spatial over-densities in specific regions of the sky due to mixing timescales in the inner Galaxy (dynamical timescale at a solar radius of $\sim250$ Myr). 
Nevertheless, several studies have shown that stars accreted a long time ago, in the first billion years after the formation of our Galaxy, could still be detected as coherent structures in kinematics-related spaces of integral of motions, which are supposed to be conserved during the merger events \citep{helmi1999, helmidezeeuw2000, knebe2005, brown2005, helmi2006, font2006, choi2007, morrison2009, gomez2010, refiorentin2015}. 
The most used kinematics-related space is the energy-angular momentum space ($E-L_z$), where E is the total energy and $L_z$ is the \textit{z} component of the angular momentum in a reference system with the $x-y$ plane set on the Galactic disc and the \textit{z} axis oriented as its spin. 

\citet{helmidezeeuw2000} showed that the stellar clumps in several of these kinematics-related spaces ($E-L_z$, $E-L$ and $L-L_z$) are still present after 12 Gyr of evolution, when the system has phase-mixed completely. Thus, the authors concluded that the space of motion integrals are the natural spaces to look for the substructures produced by accreted satellites and to infer the total number of accretion events. The Helmi Stream was the first substructure found in the MW by applying this method. \citet{helmi1999} showed that $\approx 10\%$ of the metal-poor stars ($\rm{[Fe/H]} \le -1.6$ dex) in the Galactic halo come from a single coherent structure.

These results, however, have been based on a number of assumptions, as is discussed in \citet{jean-baptiste2017}. When these assumptions are removed, accreted stars do not generally conserve their initial clumping in the so-called ‘integral (or pseudo-integral)-of-motion’ spaces (hereafter simply kinematic spaces), since none of the quantity of interest is conserved: neither the energy, nor the $z$ component of the angular momentum, nor the actions (these actions are also often used to discriminate ex situ populations \citep[see, for example, ][]{malhan22}). The limitations of this approach have been shown to affect both field stars \citep{jean-baptiste2017} and the globular cluster population \citep{pagnini23}, and have been confirmed by a number of numerical works since the early results of \citet{jean-baptiste2017} (see \citet{koppelman20, amarante22} for idealised simulations and \citet{khoperskov23a, khoperskov23b} for simulations in a cosmological context). Not only can accreted stars from the same progenitor redistribute over a large portion of the above-cited spaces, provided that the merger mass ratio is high enough ($\ge 1:10$), but in situ stars can also occupy regions of the kinematic spaces  where accreted stars (or globular clusters) are deposited \citep{jean-baptiste2017, pagnini23, khoperskov23a, khoperskov23b}, making the use of these spaces alone to reconstruct the merger history of our Galaxy questionable.

A potential way to overcome the problem and still use kinematic information to reconstruct the accretion history of the Galaxy is to make use of the metallicity distribution function (hereafter MDF) of stars located in different regions of  kinematic spaces: groups of stars that are characterised by different kinematic properties and that also have different MDFs can be indicative of independent accretion events. 
The discovery of the Sequoia merger event ($M_* \sim 5\times 10^7 M_{\odot}$) is an example of this procedure. Sequoia \citep{myeong2019} was proposed to have provided the bulk of the high-energy retrograde stars in the stellar halo \citep[already known, see][]{myeong18, koppelman18}, while only the null component ($L_z\sim0$) would be related to the \textit{Gaia}-Sausage Enceladus (hereafter GSE, $M_* \sim 6\times 10^8 M_{\odot}$) event \citep{nissenschuster2010, belokurov2018, haywood2018, helmi2018}. 
This distinction was motivated by the fact that these two structures have different MDFs: while the GSE MDF peaks at $\rm{[Fe/H]}=-1.3$, the Sequoia MDF peaks at $\rm{[Fe/H]}=-1.6$, being more metal-poor \citep{myeong2019}. A difference in [Na/Fe], [Mg/Fe] and [Ca/Fe] has also been pointed out by \citet{matsuno19}, but these abundances differ even within the various definitions of Sequoia \citep[see][]{horta23}.
The general difficulty of reconstructing clean samples for GSE and Sequoia has been rigorously addressed by \citet{feuillet20, feuillet21}. The latter study, in particular, recognises how the abundance properties of Sequoia change with its kinematic definition.
This kind of study has been applied to other regions of kinematic spaces \citep[see, for example, ][]{naidu2020, ruiz-lara22}, making use of the same overall approach: once different MDFs have been identified, the average metallicity value for each identified region of the kinematic spaces is calculated and these different average metallicities are associated with different stellar masses of the progenitor satellites by making use of mass-metallicity relations. 

Finding different average metallicities for different regions of kinematic spaces may at first glance seem proof that the general approach is correct; if different regions of kinematic spaces also have different average metallicities, the straightforward explanation is that they are populated by stellar populations of different origins. But the question is whether this approach is globally valid or whether we are using a somewhat circular argument, by making use of the results (different MDFs) to support the hypothesis (different kinematic properties identify different progenitors). It is therefore necessary to reverse the problem and ask ourselves how the fact that kinematic quantities, such as energy, angular momentum, and actions, are not generally conserved during a merger (if it is sufficiently massive) is reflected in the MDFs of stars located in different regions of kinematic spaces. It is also necessary to ask how this depends on the initial (pre-accretion) metallicity gradients of the satellite galaxies.
These questions constitute the motivations of the present work, in which we analyse N-body simulations of galaxy mergers complemented by chemical information to verify the extent to which the above described approach is robust.  \\

%\newpage 
This paper is organised as follows.
In Section \ref{sec:methods}, we describe the simulations that we use for the analysis, together with the orbital parameters and different chemical abundances gradients that we have considered. 
In Section \ref{sec:results}, we report the results in the kinematic spaces complemented by chemical information, studying in the first place the metallicity patterns in these spaces (Section \ref{sec:patterns}). Then  we  analyse MDFs in different regions of the kinematic spaces and report the results in Section \ref{sec:mdf}, varying both the volume of the simulations considered (Section \ref{sec:elzmdfsun}) and the orbital parameters (see also Appendix \ref{sec:elzmdfphi}). Furthermore, we make a tentative reconstruction of the accretion history of the Galaxy by means of these MDFs in Section \ref{sec:kirby} and analyse the relation between the metallicity gradient in the satellite and the one in the kinematic space in Section \ref{sec:relgrad}. Finally, all the conclusions of the analysis are summarised in Section \ref{sec:conclusions}.

%--------------------------------------------------------------------
\section{Numerical methods}
%--------------------------------------------------------------------
\label{sec:methods}

In this study, we analyse seven dissipationless, high-resolution N-body simulations of the accretion of a satellite onto a MW-type galaxy with different orbital parameters.
The mass ratio is 1:10 since MW-type galaxies have likely experienced such accretions during their evolution \citep[see][]{fattahi2018, fragkoudi2020}.  
These simulations are fully self-consistent, since both the satellites and the main galaxy are modelled as a collection of particles that reacts to the interaction, experiencing tidal effects and dynamical friction.
The interaction is simulated for 5 Gyr, which is sufficient time to complete the merger and have a dynamical relaxation, at least in the inner regions of the remnant galaxy. 
These simulations are the same 1:10 single accretion ($MWsat\_n1\_\Phi$) analysed in \citet{pagnini23}, which we refer to for further details.
We focus on single accretion events since we show that they already generate complex patterns that are not easily interpretable.

All the parameters are reported in Table \ref{tab:numbers}.
The total number of particles (MW and satellite) is $N_{TOT}=27 500 110$, of which $N_{MW}=25000100$ represent MW-type (stellar and dark matter) particles. 
Both galaxies consist of a thin, an intermediate, and a thick disc, mimicking the galactic thin disc, the young thick disc, and the old thick disc \citep{haywood2013, dimatteo2016}. The thin disc is as massive as the sum of the thick and intermediate ones \citep{snaith2014}. The discs have different scale heights (a) and lengths (h).
The modelled galaxies also comprise a hundred globular clusters and they are embedded in a dark matter halo. 
The discs were modelled with Miyamoto-Nagai density distributions \citep{miyamotonagai1975}:
\begin{equation}
    \rho_*(R,z)=\left(\frac{h_*^2 M_*}{4\pi}\right) 
                \frac{a_* R^2 + (a_* + 3\sqrt{z^2 + h_*^2})(a_* + \sqrt{z^2 + h_*^2})^2}
                {\left[a^2 + \left(a_* + \sqrt{z^2 + h_*^2}\right)^2 \right]^{5/2}(z^2 + h_*^2)^{3/2}}
,\end{equation}
with $M_*$, $a_*$, and $h_*$ masses and characteristic lengths and heights. 
The dark halo was modelled as a Plummer sphere:
\begin{equation}
    \rho_{halo}(r)=\left(\frac{3 M_{halo}}{4\pi a_{halo}^{3}}\right)
                   \left(1+\frac{r^2}{a_{halo}^2}\right)^{-5/2}
,\end{equation}
with $M_{halo}$ and $a_{halo}$, a characteristic mass and radius, respectively.

The number of stellar particles in the MW-like galaxy is $20\times10^6$ with a total mass of $M_*\sim9\times10^{10}M_{\odot}$; thus, it has an average particle mass of $m_*\sim4\times10^3 M_{\odot}$. The number of dark matter particles is $5\times10^6$ and the total mass is $M_{DM}\sim3.7\times10^{11}M_{\odot}$; hence, it has a particle mass of $m_{DM}\sim7\times10^4$. 

The mass and total number of particles in each disc component in the satellite is ten times smaller, while their sizes are reduced by a factor of $\sqrt{10}$ \citep[see][for the mass–size relation]{fernandez2013}, which is reported in  Table \ref{tab:numbers}. 
The total mass of the satellite has been assumed to be $M_{halo}=4.6\times10^{10}M_{\odot}$ and the visible mass $M_*=9\times10^9M_{\odot}$, which is compatible with the typical dwarf galaxies in the Local Group.
The total mass is smaller than the one estimated for GSE ($\sim10^{11}$); thus, the results found in this case would be even more evident in a GSE-like case.
The choice to use a core dark matter halo comes from a number of pieces of observational evidence that are more consistent with a nearly flat density core profile \citep{deblok2010}.

%\section{Kinematics parameters}

In a reference frame with the x-y plane initially set on the MW disc and the \textit{z} axis oriented as its spin, 
initial satellite positions and 3D velocities relative to the main galaxy are reported in Table \ref{tab:par}. Their orbital planes are inclined of $\phi_{orb}$ with respect to the MW-type galaxy disc. 

The distance of the satellite from the the MW centre is initially $D_{sat}=100$ kpc.
Initial orbital velocities of the satellite correspond to that of a parabolic orbit for a 1:10 mass ratio with a MW-type galaxy mass of 200 (in units of this mass). 
The choice of exploring parabolic orbits is in agreement with cosmological predictions \citep{khochfarburkert2006}.
The seven simulations differ in the initial orientation of the satellite orbital plane, which could have the following values: $\phi_{orb} = 0^{\circ}, 30^{\circ}, 60^{\circ}, 90^{\circ}, 120^{\circ}, 150^{\circ}, 180^{\circ}$, $\phi_{orb} = 0^{\circ}, 30^{\circ}, 60^{\circ}$ corresponding to direct orbits, $\phi_{orb} = 90^{\circ}$ to a polar orbit, and $\phi_{orb} = 120^{\circ}, 150^{\circ}, 180^{\circ}$ to retrograde orbits. 

Initial conditions were generated by adopting the iterative method described in \citet{rodionov2009}.
All simulations were run by making use of the TreeSPH code described in \citet{semelincombes2002}.
Gravitational forces were calculated using a tolerance parameter of $\theta$ = 0.7 and include terms up to the quadrupole order in the multiple expansion. A Plummer potential was used to soften gravitational forces, with constant softening lengths for different particle species. In all of the simulations described here, $\epsilon$ = 50 pc was adopted. 
The equations of motion were integrated using a leapfrog algorithm with a fixed time step of $\Delta t = 2.5 \times 10^5$ yr.
In this work, the following set of units is used: distances are given in kiloparsecs, angular momenta in $10^2$ kpc km/s, energies in $10^4$ km$^2$/s$^2$, and time in $10^7$ yr.

\begin{table}
    %\centering

    \caption{Scale lengths (a) and heights (h), mass (M), number of particles (N), and  particle mass (m) for the disc components of the MW and the satellite, considering a cut-off in radius at ten times the scale length.}
    \label{tab:numbers}
    \resizebox{\columnwidth}{!}{
    \begin{tabular}{|l|c c c c c|}
         \hline
           & a & h & M & N & m
           \\ \hline
         MW: thin disc & 4.80 & 0.25 & 16.21 & $1 \times 10^7$ & $1.6 \times 10^{-6}$\\ 
         MW: intermediate disc & 2.00 & 0.60 & 11.69 & $6 \times 10^6$ & $1.9 \times 10^{-6}$\\ 
         MW: thick disc & 2.00 & 0.80 & 8.43 & $4\times 10^6$ & $2.1 \times 10^{-6}$\\
         MW: GC system & 2.00 & 0.80 & $6.5\times 10^{-2}$ & 100 & $6.5 \times 10^{-4}$\\
         MW: dark halo & 20.00 & - & 160.00 & $5 \times 10^6$ & $3.2 \times 10^{-5}$\\
         \hline 
         Satellite: thin disc & 1.52 & 0.08 & 1.69 & $1 \times 10^6$ & $1.6 \times 10^{-6}$\\
         Satellite: intermediate disc & 0.63 & 0.19 & 1.17 & $6 \times 10^5$ & $1.9 \times 10^{-6}$\\
         Satellite: thick disc & 0.63 & 0.25 & 0.84 & $4 \times 10^5$ & $2.1 \times 10^{-6}$\\
         Satellite: GC system & 0.63 & 0.25 & $6.5\times 10^{-3}$ & 10 & $6.5 \times 10^{-4}$\\
         Satellite: dark halo & 6.32 & - & 16.00 & $5 \times 10^5$ & $3.2 \times 10^{-5}$\\
         \hline
    \end{tabular}
    }
    \tablefoot{Masses are in units of $2.3\times10^9M_{\odot}$ and distances are in kiloparsecs.}
%\end{table}
\newline
%\begin{table}
    \caption{Initial positions and velocities of the satellite.}
    \label{tab:par}
    \resizebox{\columnwidth}{!}{
    %\centering
    \begin{tabular}{|l|c c c c c c c |}
         \hline
         & $x_{sat}$ & $y_{sat}$ & $z_{sat}$ & $v_{x,sat}$ & $v_{y,sat}$ & $v_{z,sat}$ & $\phi_{orb}$  \\ \hline
         $MWsat\_n1\_\Phi$0 & 100.00 & 0.00 & 0.00 & -2.06 & 0.42 & 0.00 & 0\\ 
         $MWsat\_n1\_\Phi$30 & 86.60 & 0.00 & -50.00 & -1.78 & 0.42 & 1.03 & 30\\
         $MWsat\_n1\_\Phi$60 & 50.00 & 0.00 & -86.60 & -1.03 & 0.42 & 1.78 & 60\\
         $MWsat\_n1\_\Phi$90 & 0.00 & 0.00 & -100.00 & 0.00 & 0.42 & 2.06 & 90\\
         $MWsat\_n1\_\Phi$120 & -50.00 & 0.00 & -86.60 & 1.03 & 0.42 & 1.78 & 120\\
         $MWsat\_n1\_\Phi$150 & -86.60 & 0.00 & -50.00 & 1.78 & 0.42 & 1.03 & 150\\
         $MWsat\_n1\_\Phi$180 & -100 & 0.00 & 0.00  & 2.06 & 0.42 & 0.00 & 180\\
         \hline
    \end{tabular}
    }
    \tablefoot{The angle, $\phi_{orb}$, is the inclination of the satellite orbital plane with respect to the MW disc (in degrees). Distances are in kiloparsecs and velocities are in units of 100 km/s.}
\end{table}

%\newpage
%\subsection{Assigning chemical abundances to the simulated galaxies}
%\label{sec:chem}

As was stated in the previous section, the simulations analysed in this paper are dissipationless, which implies that no star formation or chemical enrichment was modelled. However, following an approach already used in a number of works \citep[see, for example, ][]{dimatteo13, martinez13, fragkoudi17, fragkoudi18, khoperskov18}, we have assigned, a posteriori, chemical abundances to the stellar particles of the MW-type galaxy and the satellite based on their kinematic properties and the current observational constraints, as is detailed below. 

\paragraph{The case of initial vertical abundance gradients}
In this first case, the mean metallicity of disc particles is independent of radius; that is, no radial metallicity gradient is initially present in either the disc of the MW-type galaxy or in the satellite. 
However, a vertical abundance gradient has been taken into account on the basis of observational evidence in the MW. In our Galaxy, indeed, the thick disc is on average more metal-poor and more $\alpha-$enhanced with respect to the thin disc \citep{fuhrmann1998, haywood2013, bensby2014, bovy2012a}. In particular, in the inner regions of the Galaxy, where all the disc components are present, the metallicity and the $\alpha$ abundance can be described by different Gaussian distributions \citep{hayden2015}.

For the MW-type galaxy, then, the [Fe/H] and the [Mg/Fe] values were assigned to the particles with normal distributions, with three different mean ($\mu$) and standard deviation ($\sigma$) values for the thin, intermediate, and thick discs, with decreasing mean values for the mean metallicity and increasing ones for the $\alpha$ abundances (see Table \ref{tab:nogradMW}).\\

\begin{figure}
%    \resizebox{\columnwidth}{!}{
    \centering
    \includegraphics[width=1\linewidth]{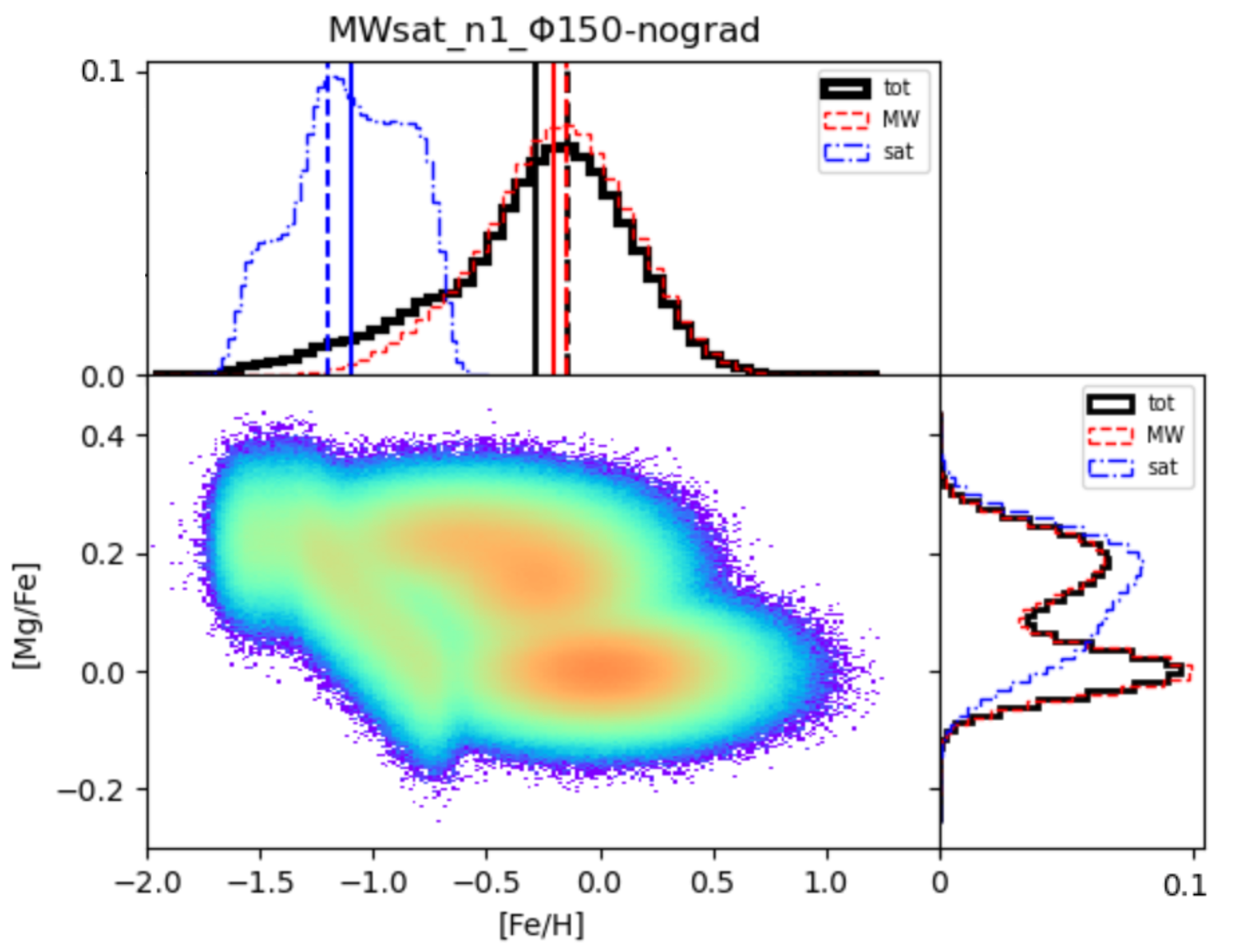}
    \\
    \includegraphics[width=1\linewidth]{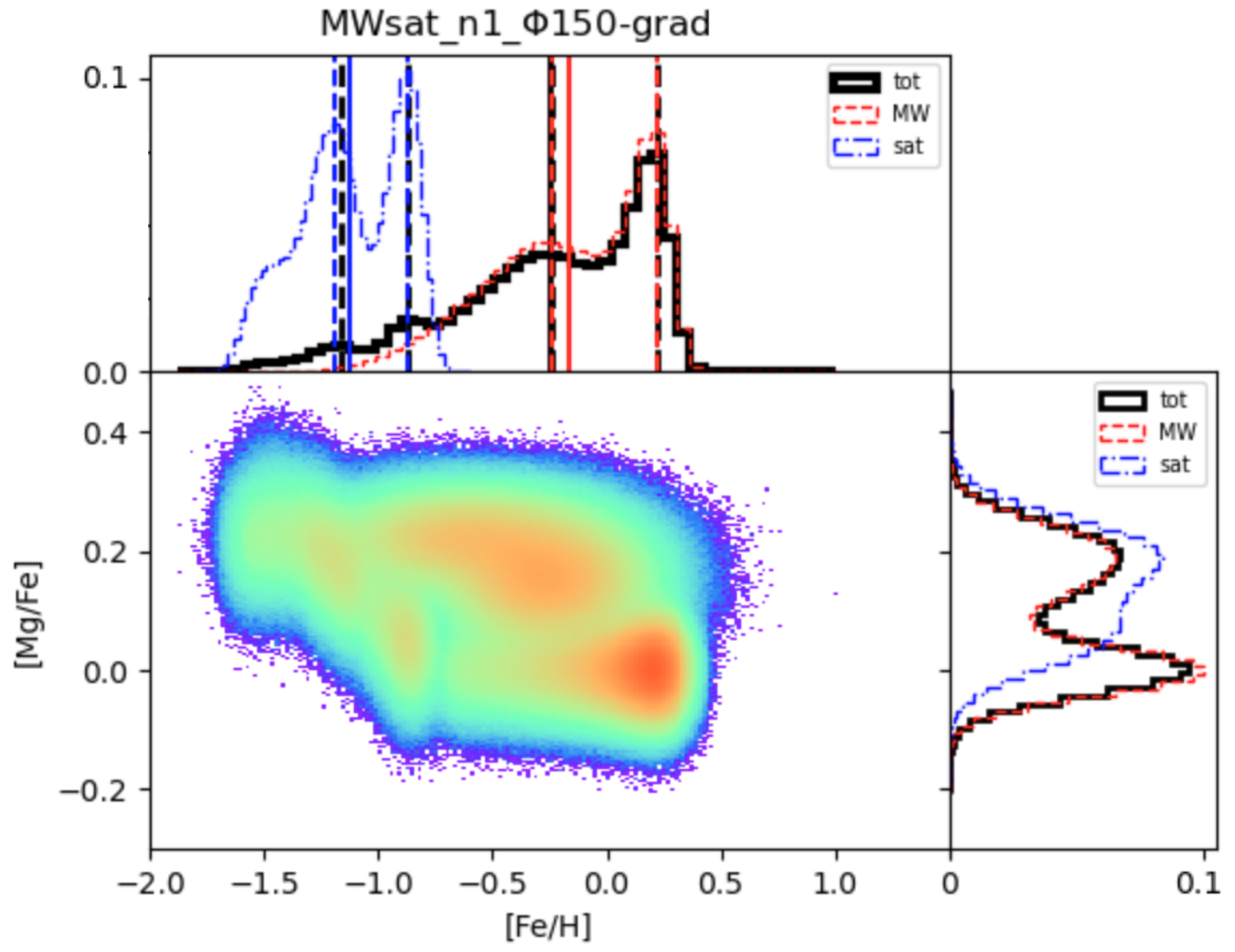}
%    }
    \caption{The $\rm{[Mg/Fe]}-\rm{[Fe/H]}$ space and its normalised marginal distributions, in the case of a vertical metallicity gradient only (upper panel) and in the case of an additional radial metallicity gradient (lower panel).}
    \label{fig:chem}
\end{figure}

As for the satellite galaxy, uniform distributions between different minimum and maximum values were used to assign metallicities to its thin, intermediate, and thick discs (Table \ref{tab:nogradsat}), to which a dispersion of 0.05 dex was added. In the first place, we assumed a uniform distribution for the sake of simplicity. This simple hypothesis, indeed, already generates a great complexity in the final metallicity patterns in the E-Lz space, and thus is enough for the point we wanted to make out.
The choice to assign different metallicity intervals to the satellite disc components simply mimics the case in which satellite stars with hotter kinematics also have lower mean metallicities.  
On average, the mean metallicity of the satellite is lower than that of the MW-type galaxy, as was found in observational studies \citep{kirby2013}. 
As for the [Mg/Fe] abundances of the satellite galaxy, we assigned them in such a way to reproduce qualitatively the behaviour of $\alpha-$elements with [Fe/H], as was observed in a number of dwarf galaxies of the Local group \citep[see, for example, ][]{matteucci2007, tolstoy2009}. These low-mass systems tend to show a nearly flat dependence of $\alpha$ elements with [Fe/H] until a characteristic value of [Fe/H]  (the so-called ‘knee’) at which the [$\alpha$/Fe] ratio shows an inflection. For metallicities larger than the value at the knee (hereafter [Fe/H]$_{knee}$), the [$\alpha/Fe$] ratios show a decrease with increasing [Fe/H]. The value of [Fe/H] at the knee of the [$\alpha$/Fe]--[Fe/H] relation is in general not the same for all dwarf galaxies \citep[see, for example, Fig.~11 in][]{tolstoy2009} but changes from galaxy to galaxy, reflecting their different star formation histories. To mimic this behaviour, we assigned [Mg/Fe] ratios to the satellite stellar particles in the following way: 
\begin{align}
\rm{[Mg/Fe]}_{sat} &=
  \begin{cases}
    \rm{[Mg/Fe]}_{knee} & \text{if } \rm{[Fe/H]}_{sat}<  \rm{[Fe/H]}_{knee},\\
\rm{[Fe/H]}_{sat}*m +q & \text{if } \rm{[Fe/H]}_{sat}\geq \rm{[Fe/H]}_{knee}
  \end{cases}
  \end{align}
with $\rm{[Mg/Fe]}_{knee}=0.22$, $\rm{[Fe/H]}_{knee}=-1.3$, $m = - 0.4$, and $q = - 0.3$. With such a choice of values, the distribution of in situ and accreted stars in the [Mg/Fe]--[Fe/H] plane is qualitatively similar\footnote{We emphasise that the chemical abundances sequence generated with our approach are not a fit to any data, but are simply intended  to capture the main trends found in observational [Mg/Fe]--[Fe/H] planes.} to that found for stars in the solar vicinity \citep{nissenschuster2010} or in a volume a few kiloparsecs from the Sun \citep{hayes2018} (see Fig.~\ref{fig:chem}, top panel). In particular, the choice of $\rm{[Fe/H]}_{knee}=-1.3$ was made to be in agreement with the value of metallicity at which \citet{nissenschuster2010} have reported the appearence of the low-$\alpha$, accreted, halo sequence.

\begin{table}%[h]
    \centering
    \caption{Mean and standard deviation values of the normal distributions that define the [Fe/H] and [Mg/Fe] values for the particles of the thin, intermediate, and thick discs of the MW-type galaxy.}  
    \label{tab:nogradMW}
    \begin{tabular}{|l|c c|}
         \hline
         & $\mu$ & $\sigma$ \\\hline
         $\rm{[Fe/H]}_{MW,thin}$ & 0.0 & 0.25\\
         $\rm{[Fe/H]}_{MW,inter}$ & -0.26 & 0.2\\
         $\rm{[Fe/H]}_{MW,thick}$ & -0.62 & 0.26\\
         \hline
         $\rm{[Mg/Fe]}_{MW,thin}$ & 0.0 & 0.04\\
         $\rm{[Mg/Fe]}_{MW,inter}$ & 0.15 & 0.05\\
         $\rm{[Mg/Fe]}_{MW,thick}$ & 0.22 & 0.04\\
         \hline
    \end{tabular}
\end{table}

\begin{table}%[h]
    \centering
    \caption{Minimum and maximum values of the uniform distributions used to assign a metallicity value to the satellite particles.}
    \label{tab:nogradsat}
    \begin{tabular}{|l|c c|}
         \hline
         & min & max \\\hline
         $\rm{[Fe/H]}_{sat,thin}$ & -1.1 & -0.7\\
         $\rm{[Fe/H]}_{sat,inter}$ & -1.3 & -1.1\\
         $\rm{[Fe/H]}_{sat,thick}$ & -1.6 & -1.3\\
         \hline
    \end{tabular}
\end{table}

\paragraph{Adding an initial radial metallicity gradient}

In this second case, we have explored the possibility that both the thin discs of the MW-type galaxy and the satellite initially show a negative radial metallicity gradient that co-exists with the vertical gradients discussed before.

In the thin disc of the MW, indeed, it is known that the inner regions have $\rm{[Fe/H]}\sim0.3$, while the outer regions (beyond 10~kpc) have $\rm{[Fe/H]}\sim-0.7$ \citep[e.g.][]{hayden2015}. 
A metallicity gradient is currently absent in the thick disc \citep[][]{cheng2012, hayden2015}, a fact that could be due to its formation during a more turbulent phase, which would have led to an efficient mixing of metals \citep[][]{haywood2013, haywood15, lenhert14}. 
The assignment of the radial metallicity gradient to the stellar particles was done on the basis of their initial ($t=0$) positions.\\
In particular, the metallicity of the thin disc particles of the MW-type galaxy was assigned as follows:
\begin{equation}\label{eqn_radgrad_MW_Rmax}
    \rm{[Fe/H]}_{MW,thin} = \rm{[Fe/H]}_{MW,thin}^{max} - \frac{R}{R_{max,MWthin}}
,\end{equation}
where $R$ is the  distance of the particle from the MW-type galaxy centre, projected in the disc plane, $\rm{[Fe/H]}_{MW,thin}^{max}$ is the maximum metallicity at the centre of the thin disc ($R=0$), which we assigned to be equal to 0.3~dex, and $R_{max,MWthin}$ is the maximum distance from the galaxy centre at which stellar particles of the thin disc of the MW-type galaxy are found. With $R_{max,MWthin}$ being equal to 50~kpc, this implies that Eqn.~\ref{eqn_radgrad_MW_Rmax} can be rewritten as
\begin{equation}\label{eqn_radgrad_MW}
    \rm{[Fe/H]}_{MW,thin} = \rm{[Fe/H]}_{MW,thin}^{max} - f_{MWthin}*R
,\end{equation}
with $f_{MWthin}=0.02$ dex/kpc. We caution the reader that the metallicity gradient of the MW disc is known to be more complex than what we have used, with a steeper slope a few kiloparsecs from the Sun \citep[see][for a recent study]{haywood24}. %et al 2023, A\&A submitted, for a recent study). 
While the adopted values of the gradient can change the mean metallicities of stellar populations and the shapes of the MDFs, the bottom line of the results that we present below does not change.

Similarly, for the thin disc of the satellite galaxy,
\begin{equation}
   \rm{[Fe/H]}_{satthin} = \rm{[Fe/H]}_{satthin}^{max} - f_{satthin}*R
,\end{equation}
where $R$ is now the distance of a stellar particle of the satellite from the satellite centre, $\rm{[Fe/H]}_{sat,thin}^{max} $ is the maximum metallicity at the centre of the satellite thin disc (which we assigned to be equal to -0.8~dex), and  $f_{satthin}=0.05$~dex/kpc, in such a way that the satellite thin disc has a metallicity of -1.5~dex at its edge. This gradient is consistent with the ones observed in Local Group dwarf galaxies \citep[e.g., see][]{taibi22}. In cosmological simulations, similar gradients are also observed \citep[e.g.,][]{khoperskov23c}.

The bottom panel of Figure \ref{fig:chem} shows the distribution of stars (accreted and in situ) in the $\rm{[Mg/Fe]}-\rm{[Fe/H]}$ space when an initial radial metallicity gradient is added. The main trends are those observed already in the case of a vertical gradient only: the flatness of the [Mg/Fe] ratio for [Fe/H]$\lesssim -1.3$~dex, the appearance of a low-$\alpha$ sequence at higher metallicities, and the redistribution of stars into an $\alpha$-enhanced thick disc and a low-$\alpha$ thin disc counterpart. To these general features, the inclusion of a radial metallicity gradient has the effect of introducing a certain amount of complexity: the low-$\alpha$ thin disc sequence and the low-$\alpha$ accreted sequence are no longer uniformly populated, but show prominent metal-rich peaks that clearly appear in the total MDF, as well as in the MDFs of the satellite and of the MW-type galaxy, taken separately. The $\alpha-$DFs, instead, stay the same, since in this second scenario no hypothesis about an $\alpha$-gradient with radius was taken into account. In particular, in both cases (with or without a radial metallicity gradient) the $\alpha-$DF of the MW-type galaxies is bimodal, as has been found for stars in the MW disc \citep[][]{haywood18}.  \\

In the following, we refer to the two different initial conditions for the metallicity gradients with the suffix
‘{\it nograd}’, to indicate the case in which no radial metallicity gradient is initially present in the main and satellite discs, and with the suffix ‘{\it grad}’, to indicate the case in which radial metallicity gradients are initially present. 
This suffix was added to the identificator of the simulation, which was defined in the previous section. For example, the simulation named MWsat\_n1\_$\Phi$0\_grad refers to a single 1:10 merger, in which the satellite is initially on a planar prograde orbit and in which the thin discs of both the MW-type galaxy and the satellite are initially characterised by a radial metallicity gradient.

%--------------------------------------------------------------------
\section{Results}
%--------------------------------------------------------------------
\label{sec:results}
%kinematics results
As far as the kinematics-related spaces are concerned, we confirm earlier results \citep{jean-baptiste2017, koppelman20, panithanpaisal21, amarante22, khoperskov23b}; that is, energy and angular momentum are not generally conserved quantities for such a significantly 
massive satellite, and this is independent of the specific orbital inclination of the satellite galaxy relative to the main galaxy disc. 
Therefore, a single accretion event can redistribute stars over a large extent of the $E-L_z$ space and these stars are not smoothly redistributed in this space. They form clumps whose number and density depend on the number of passages the satellite has experienced around the MW-type galaxy before complete coalescence and on the mass loss it has experienced at each passage. The stars lost by the satellite in the early phases of its accretion onto the Galaxy tend to redistribute in the upper part of the $E-L_z$ space; that is, at high energies. These stars also span a wide range of $L_z$ values. Hence, one accreted satellite gives rise to several overdensities in the $E-L_z$ space. On the other hand, stars which were in the disc of the main galaxy before the accretion are kinematically heated during the interaction, acquiring halo-like kinematics, and thus contributing to the formation of a stellar halo together with the accreted stars \citep[][and references therein]{zolotov09, purcell10, dimatteo2019}. As a consequence, they redistribute in $E-L_z$ space in a region that partially overlaps with that of accreted stars.

\begin{figure*}[h]
%    \resizebox{\columnwidth}{!}{
%    \centering
    \includegraphics[width=1.01\linewidth]{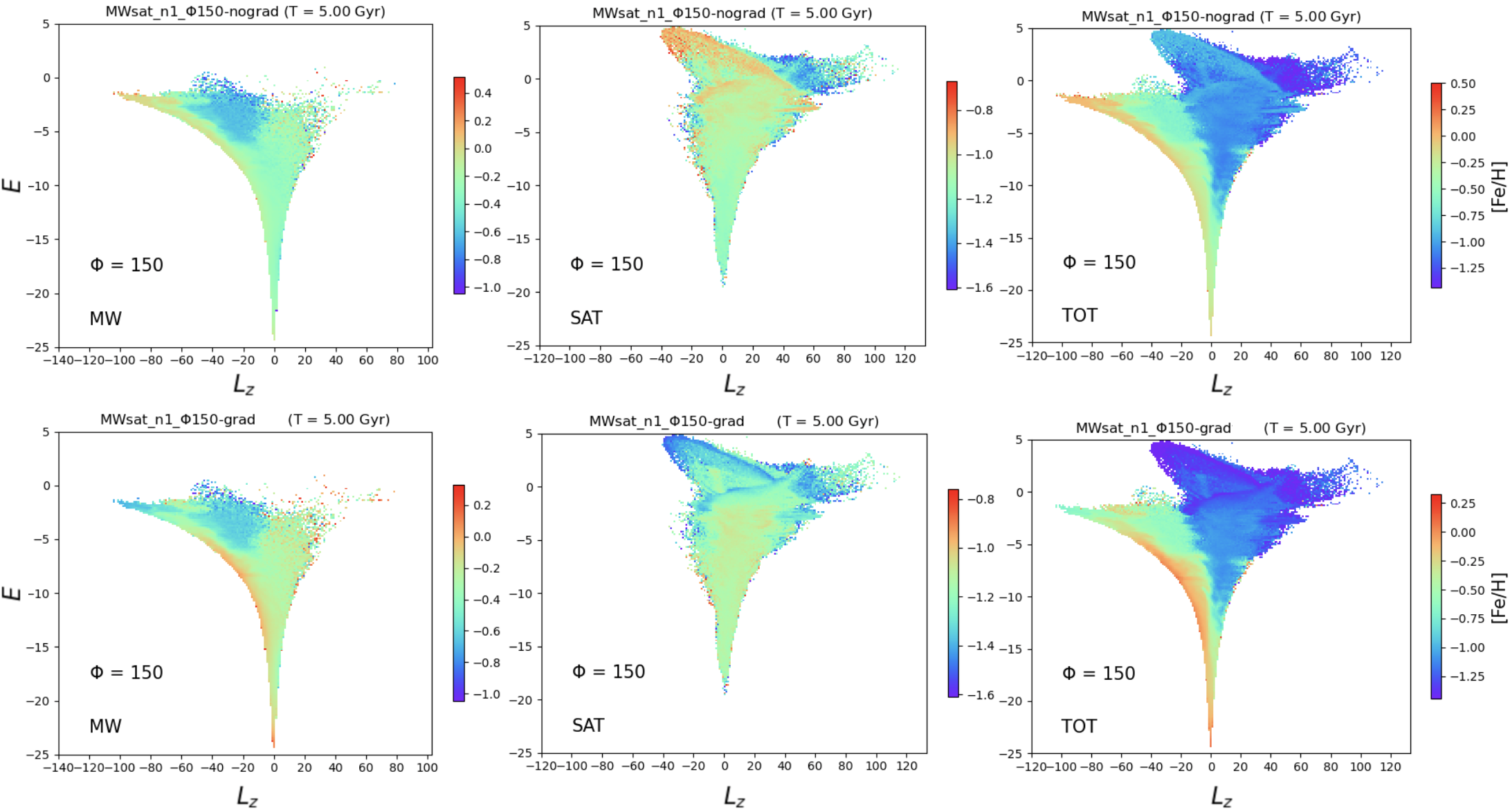}
%    }
    \caption{
    $E-L_z$ space of in situ (first column), satellite (second column), and total (third column) star particles in the simulation, colour-coded at each point with the mean metallicity. The first and second rows show the $E-L_z$ space in the case of an initial vertical gradient only and in the case of an additional initial radial metallicity gradient, respectively. The scales of the colour bars are different depending on which galaxy is considered, in order to make their specific metallicity patterns more evident. We also recall that in these plots stars on prograde orbits (which rotate as the disc) have negative $L_z$, while stars on retrograde orbits have positive $L_z$.
    }
    \label{fig:elz}
\end{figure*}

%\newpage
\subsection{Metallicity patterns in kinematic spaces}
\label{sec:patterns}

After assigning the metallicity to the star particles of the simulation, the first step we took was to study the kinematics-related spaces with this additional information, in order to see where stars with different metallicities would redistribute and to check for a dependence on differences in the initial conditions of the metallicity gradients.
We started by discussing the simulation with $\phi_{orb}=150^{\circ}$ to show the general trends in the kinematics-related spaces, then we considered different volumes of the simulation to verify the feasibility of this kind of analysis with local observational data. Finally, we compared it with the outcomes of the other simulations with different orbital parameters. \\

In Figure \ref{fig:elz}, we show the distributions in the $E-L_z$ space, since among the different kinematics-related spaces it is the one most used in the literature. They are colour-coded at each point by the mean metallicity of all the overlapping star particles. 
The different columns of Figure \ref{fig:elz} correspond to MW-type galaxy (first column), satellite (second one), and total (third one) stellar particles of the simulation. The first row shows the distributions in the $E-L_z$ space of all the star particles of the simulation in the ‘nograd’ case, resulting in both thin discs being more metal-rich than the thick ones on average. The second row shows the $E-L_z$ of the stars in the case of an additional initial radial metallicity gradient. 
The scales of the colour bars are different depending on which galaxy is considered, in order to make their specific metallicity patterns emerge in a more evident way. For the same purpose, the metallicity colour bar does not cover the whole metallicity range of the metallicity distribution of the star particles of the galaxies of the simulation, but we have considered the percentiles of the distribution and restricted the range of the colour bar to $98\%$ of the metallicity distribution, as is done in the rest of this analysis.

As far as the in situ star distribution in the $E-L_z$ space is concerned (top left plot), we can observe where the different disc components of the MW-type galaxy, characterised by different mean metallicities, dominate in the $E-L_z$ space. We can see that the most metal-rich stars (in orange-red) redistribute in the region of prograde circular orbits of the thin disc, while the most metal-poor stars (violet-blue) are mainly found in the $E-L_z$ region dominated by the thick disc (see Appendix \ref{sec:discs} for the distributions of the thin, intermediate, and thick disc stars in the $E-L_z$ space separately).\\
\begin{figure}
    \includegraphics[width=1.015\linewidth]{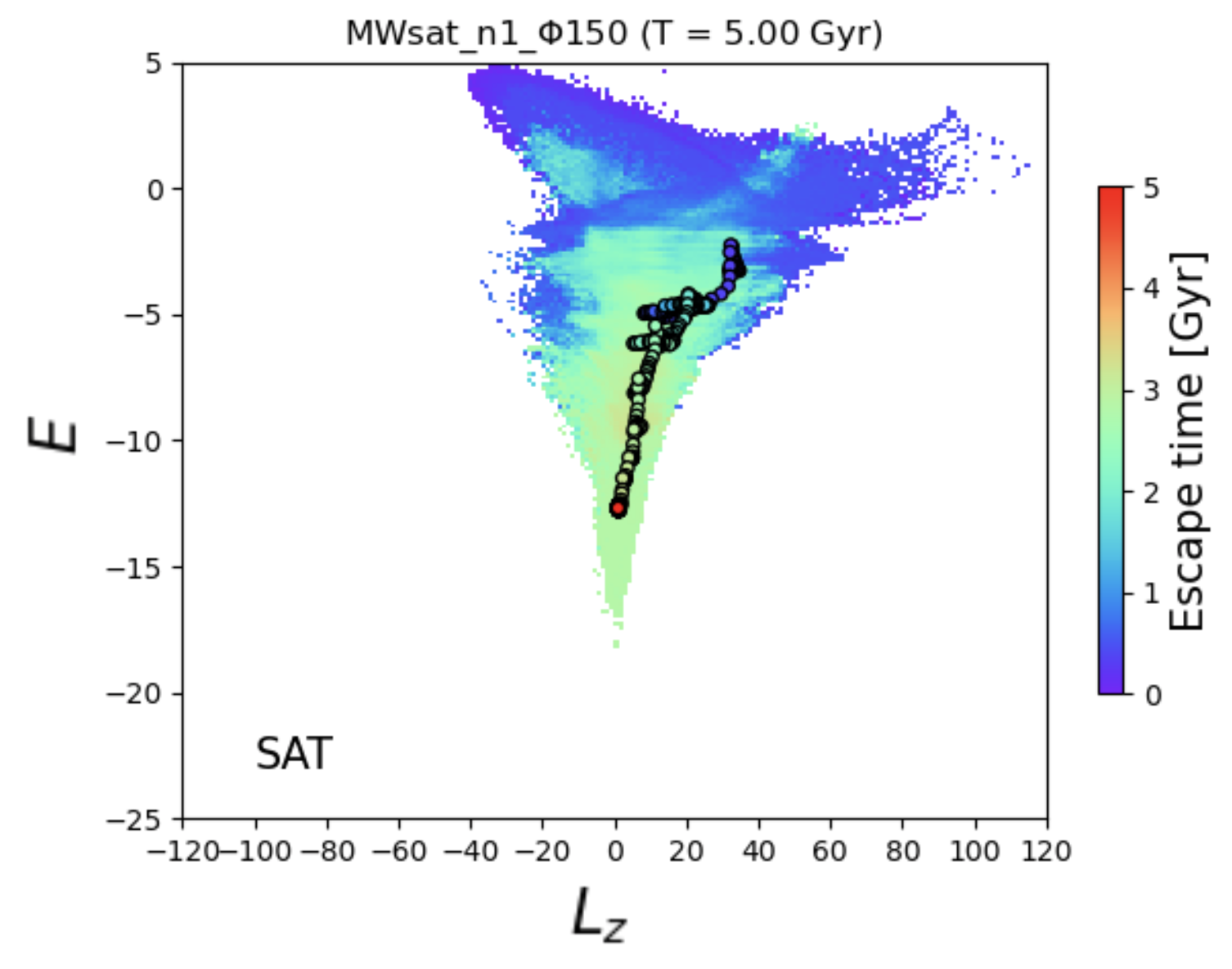}
    \caption{Final distribution of the satellite stars in the $E-L_z$ space as in the middle column of Figure \ref{fig:elz}, but colour-coded by their escape time from the satellite potential. The dots represent the mean values of energy and z component of the angular momentum of bound satellite stars, defined by being less than 15 kpc from its barycentre. There are five hundred dots (one for each snapshot of the simulation) and they are colour-coded with the same colour-bar scale but for the time of the simulation (every 10 Myr to cover the 5 Gyr time interval).}
    \label{fig:esctime}
\end{figure}

Regarding the satellite stars' distribution (top middle plot), instead we can notice that there is a region in the upper part of the $E-L_z$ space where the mean metallicity is higher on average. This region is characterised by very high values of energy, and thus is the region populated by the first stars that escape from the satellite during its interaction with the main galaxy, getting bound to the MW-type galaxy potential (Figure \ref{fig:esctime}).
These stars are those on the outskirts of the satellite thin disc; indeed, since they are at larger (radial) distances from the satellite baricenter, they are less bound by its potential, and thus more inclined to leave it. In the case we are considering now of no radial gradient in the thin disc, but only a negative vertical metallicity gradient, the thin disc stars are the most metal-rich, which explains the mean metallicity pattern of the satellite distribution in the $E-L_z$ space. 

%\newpage
Thus, the existence of an initial metallicity gradient inside the satellite has a consequence in the $E-L_z$ space: different regions show different mean metallicities. We check in the following Section \ref{sec:mdf} how this fact is reflected in the MDFs of the different regions, too, in order to see if in addition to a change in the mean value of the distribution, its shape also changes.
Finally, as far as the total star distribution is concerned (top right plot), we can observe a strong metallicity gradient in the $E-L_z$ space, which reflects the regions where the more metal-rich in situ stars dominate on more prograde orbits and have lower energy values and the regions where instead most of the more metal-poor satellite stars lie on more retrograde orbits and have higher energy values. 
In addition to the fact that on average in situ stars are more metal-rich than satellite ones, we can still observe that inside the distribution of these two galaxies there are metallicity gradients, as was described above.\\

In order to verify whether these results would be dependent on differences in the initial conditions for the metallicity gradients, we have also studied the case of an {\it additional radial negative gradient in the thin discs of both the satellite and the MW-type galaxy}, and the outskirts of the discs  thus being more metal-poor than the inner parts. 
We report the corresponding results in the second row of Figure \ref{fig:elz}: the MW-type galaxy distribution (left plot) overall resembles the one seen previously in the case of the vertical metallicity gradient only. However, we can notice the effect of having introduced a negative radial gradient too:  the region of the $E-L_z$ space populated by the thin disc (towards the edge of the $E-L_z$ distribution, for negative values of $L_z$) is not homogeneously coloured anymore, as it was in the case of the vertical metallicity gradient only, but exhibits a gradient. We can see that the upper regions are more metal-poor than in the previous case. These regions at high energies are indeed populated by  stars that initially are less gravitationally bound (i.e. high energies and large distances from the main galaxy centre). 
On the other hand, the low-energy regions are more metal-rich than before, because the populations that are initially more gravitationally bound (and more metal-rich) are also closer to the galaxy centres.
As far as the satellite distribution is concerned (middle plot), we can again observe the consequence of the choice made for the initial metallicity gradient: the upper region that was populated before by the most metal-rich stars is now on the contrary populated by the most metal-poor ones. This is clearly a consequence of assigning a radial metallicity gradient to satellite disc stars. 
Finally, regarding the total distribution (right plot), we can observe an overall metallicity gradient again, which is even stronger than before (see Section \ref{sec:relgrad}), and we can still see the gradients inside the two galaxies. \\

\subsubsection{Solar-like volume}
\label{sec:patternssun}

\begin{figure*}%[htp]
    \includegraphics[width=1.00\linewidth]{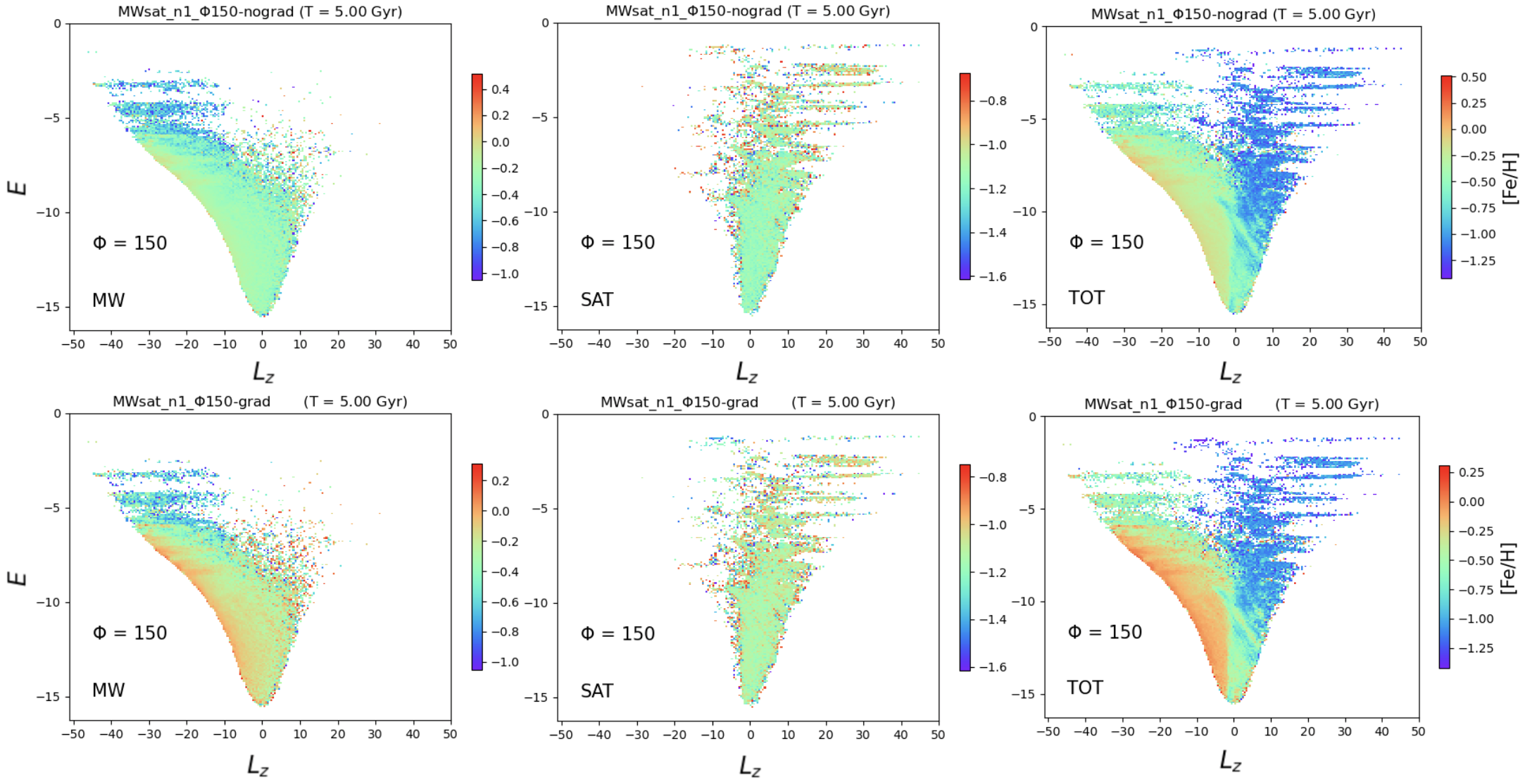}
    \caption{
    $E-L_z$ space of in situ (first column), satellite (second column), and total (third column) star particles in the simulation, colour-coded at each point with the mean metallicity, when the analysis is restricted to a solar-like volume. The first row shows the $E-L_z$ space in the case of an initial vertical metallicity gradient only. Likewise, the second row shows it, but in the case of an additional initial radial metallicity gradient. The scales of the colour bars are different depending on which galaxy is considered, in order to make their specific metallicity patterns emerge in a more evident way. We recall that in these plots stars on prograde orbits (which rotate as the disc) have negative $L_z$, while stars on retrograde orbits have positive $L_z$.
    }
    \label{fig:elzsun}
\end{figure*}

 Once the previous results for the total distributions of star particles in the $E-L_z$ space were obtained, we were interested in understanding what it was actually possible to obtain with the data we had in a limited volume observed. Thus, in order to check the feasibility of detecting accreted structures in this space in local volumes, we restricted our analysis to a ‘solar-vicinity’ or a ‘solar-like volume’. It was defined as a spherical volume that is centred at 8 kpc from the MW-type galaxy centre and that has a radius of 5 kpc, which is the typical size around the Sun for which we currently have accurate-enough parallaxes, proper motions, and line-of-sight velocities. In order to take into account the fact that accreted stars do not necessarily redistribute uniformly, we considered several different volumes homogeneously distributed in the azimuth position of the centre: $\phi_{\odot}=0^{\circ}, 90^{\circ}, 180^{\circ}, 270^{\circ}$ (in cylindrical coordinates). In the following analysis, we show the results for the first solar volume (with $\phi_{\odot}=0^{\circ}$), which are also valid for all the other volumes examined. Figure \ref{fig:elzsun} shows the final distributions in the $E-L_z$ space when the solar-like volume is considered, colour-coded at each point with the mean metallicity.

We can see in the right-most panels of Figure \ref{fig:elzsun} that the same overall trends observed for the whole volume are found: a metal-poor population with retrograde or null $L_z$, and a more metal-rich one as the orbits become more prograde (negative $L_z$), in both the metallicity gradient cases. 
Furthermore, it is interesting to notice how a retrograde single-merger event can produce a Sequoia-like debris, which corresponds to the outer parts of its GSE-like progenitor.
However, when the analysis is restricted to 5 kpc around the Sun, we can no longer capture the very metal-rich (metal-poor) satellite stars found at very high energies (see, for example, the orange (blue) color in the middle panel of the top (bottom) row of Figure \ref{fig:elz}).

This implies that the MDF of the GSE may still exhibit biases, particularly in the absence of observations for the most metal-rich (metal-poor) stars \citep[consistently with the results of][]{carrillo24}. This limitation could potentially be addressed more comprehensively with upcoming surveys such as SDSSV/4MOST (allowing us to get a more complete sample of the accreted stars now bound to the MW, but outside the solar-like volume we defined) or, more promisingly, with MOONS. MOONS will survey the bulge of the Galaxy \citep[][]{gonzalez20}, allowing us to explore the eventual accreted populations in the innermost regions of the MW and check for differences in the metallicity distributions with respect to those in the solar neighbourhood.

%\newpage
\subsubsection{Exploring different orbital inclinations}
\label{sec:patternsphi}

\begin{figure*}%[H]
%    \resizebox{\columnwidth}{!}{
    \centering
    \includegraphics[width=0.8\linewidth]{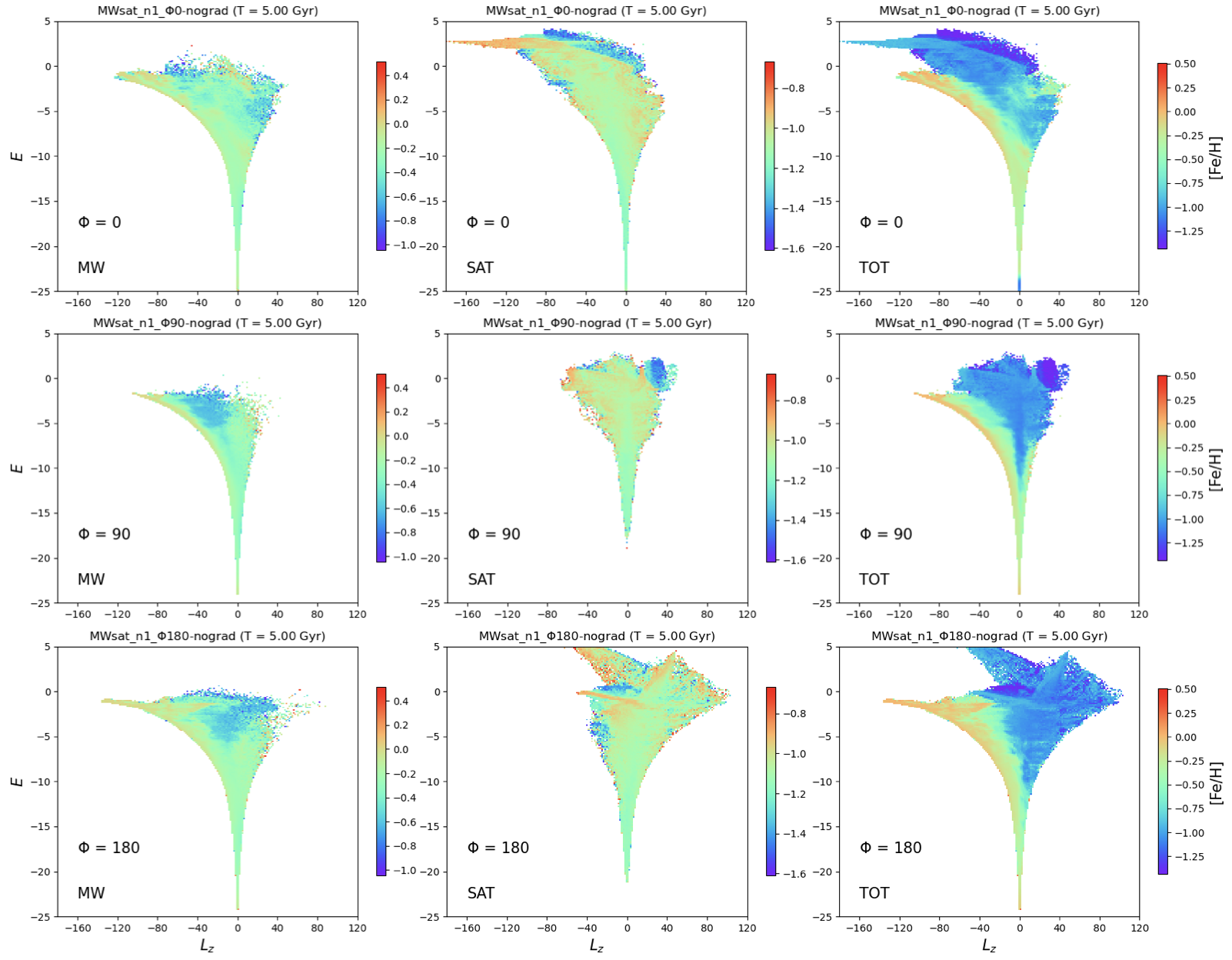}%, height=3.8cm]
    \vspace{1cm}
%    }
%    \caption{$E-L_z$ spaces of the in-situ (first \textbf{column}), satellite (second \textbf{column}) and total (third \textbf{column}) star particles in the simulation, colour-coded for their mean metallicity, for the case of an initial vertical metallicity gradient only. Each \textbf{row} shows a simulation with a different value of the initial inclination of the satellite orbital plane: $\phi_{orb}=0^{\circ}, 90^{\circ}, 180^{\circ}$.}
    %\label{fig:elzphing}
%\end{figure*}
%\begin{figure*}%[H]
%    \resizebox{\columnwidth}{!}{
%    \centering
    \includegraphics[width=0.8\linewidth]{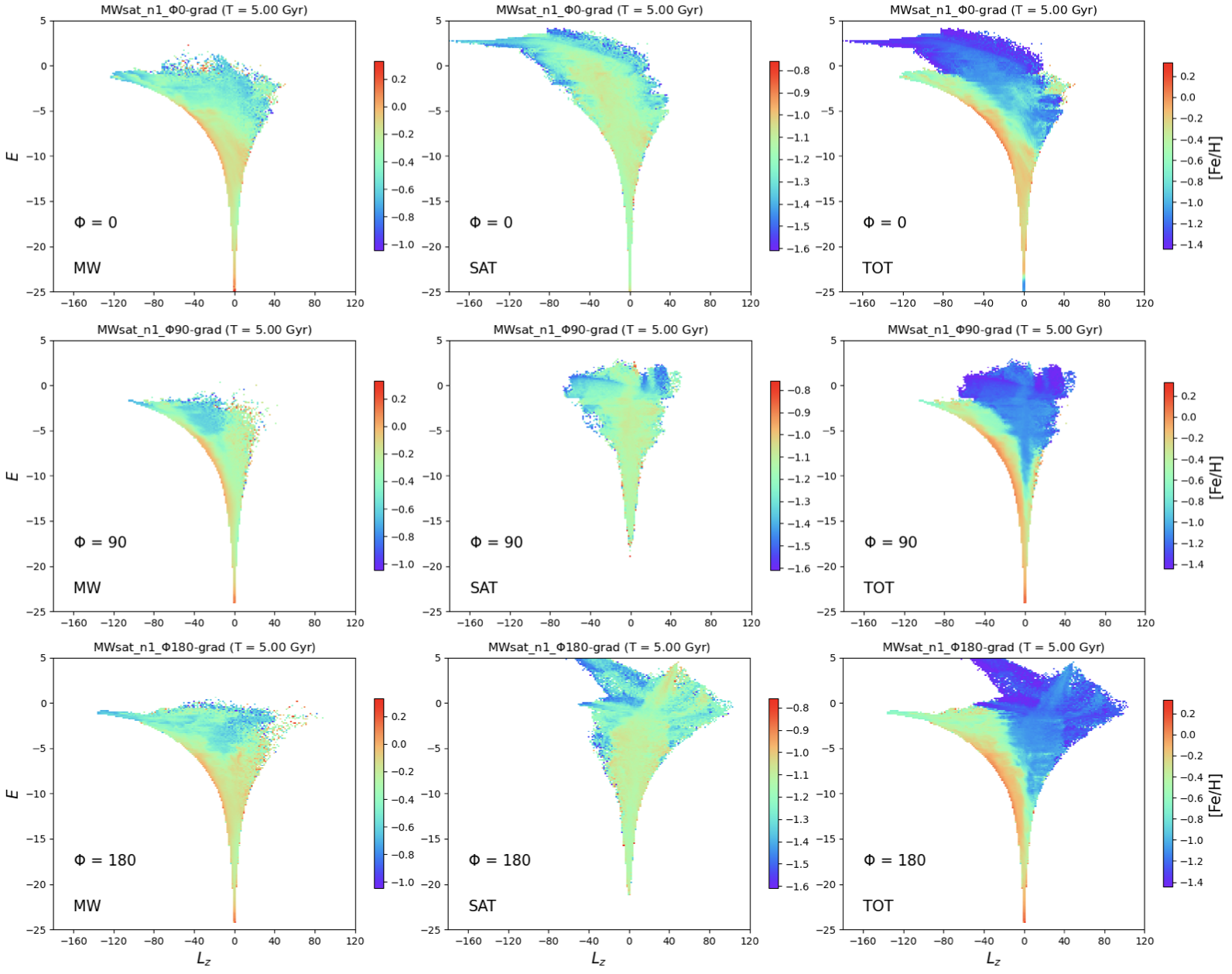}
%    }
    \caption{$E-L_z$ spaces of the in situ (first column), satellite (second column), and total (third column) star particles in the simulation, colour-coded at each point with the mean metallicity. Each row shows a simulation with a different value of the initial inclination of the satellite orbital plane: $\phi_{orb}=0^{\circ}, 90^{\circ}, 180^{\circ}$. The upper (lower) panel is for nograd (grad) case.
    %the case of an initial vertical metallicity gradient only
    %Same as Figure \ref{fig:elzphing}, but for the case of an additional initial radial metallicity gradient.
    }
    \label{fig:elzphi}
\end{figure*}

\begin{figure*}%[h!]
    \includegraphics[width=0.51\linewidth]{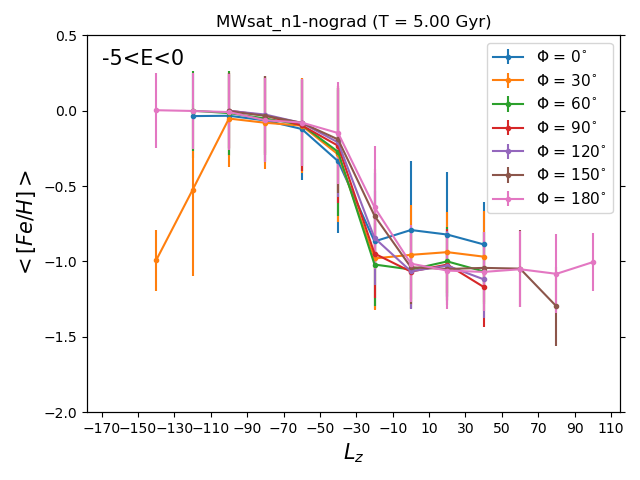} % 
    \includegraphics[width=0.51\linewidth]{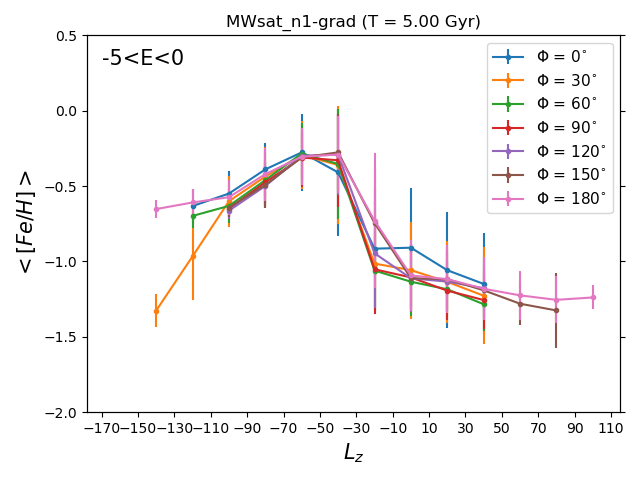}  \\  %
    \includegraphics[width=0.51\linewidth]{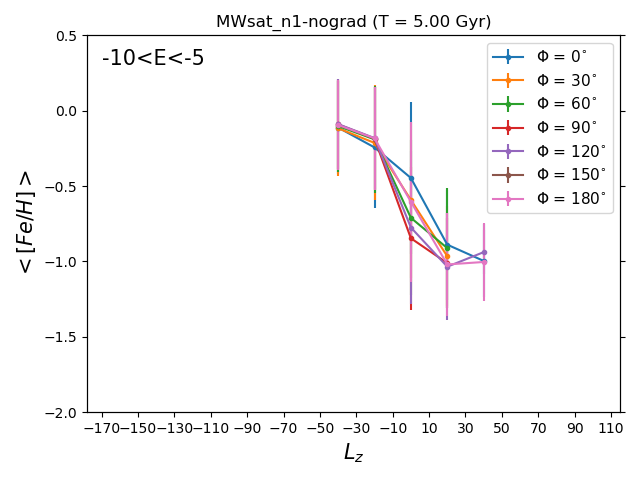} %  
    \includegraphics[width=0.51\linewidth]{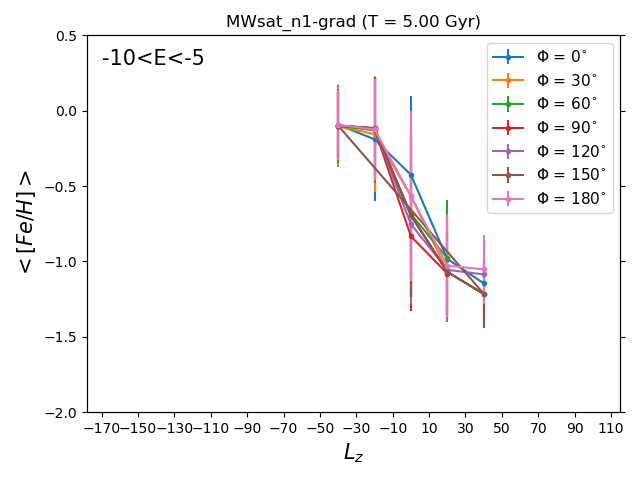}
    
    \caption{Mean metallicity and standard deviation as a function of the angular momentum interval of the $E-L_z$ space (in every column of Figures \ref{fig:elzgrid}, \ref{fig:elzgrids}) for all the simulations with $\Phi=0^{\circ}, 30^{\circ}, 60^{\circ}, 90^{\circ}, 120^{\circ}, 150^{\circ}, 180^{\circ}$, in the energy range $-5<E<0$. The left panel shows the case with the initial vertical metallicity gradient only, while the right panel shows the one with the additional radial metallicity gradient.}
    \label{fig:MDFmeanphiEall}
\end{figure*}

In order to check the dependence of our results on the satellite orbital parameters, we analysed the star distributions in all the previous spaces (colour-coded at each point with the mean metallicity) for all the other simulations with a different initial inclination of the satellite orbital plane with respect to the MW-type galaxy, $\phi_{orb}=0^{\circ}, 30^{\circ}, 60^{\circ}, 90^{\circ}, 120^{\circ}, 150^{\circ}, 180^{\circ}$.
To seek a synthesis, in the following we only report the results for the $E-L_z$ space for $\phi_{orb}=0^{\circ}, 90^{\circ}, 180^{\circ}$. %\\
In Figure \ref{fig:elzphi}, 
we show the final distributions in the $E-L_z$ space colour-coded at each point with the mean metallicity for the total volume of the simulations, with $\phi_{orb}=0^{\circ}, 90^{\circ}, 180^{\circ}$ in different rows, for the in situ (first column), satellite (second column), and total (third column) star particles. %\\

The top panel of Figure \ref{fig:elzphi} shows the distributions in the case of an initial vertical metallicity gradient only.
The in situ distributions are all very similar. 
The satellite distributions, instead, show a dependence on the orbital parameter, $\phi_{orb}$, with: (i) more metal-poor stars on prograde orbits (negative values of $L_z$) at high energies for initial prograde orbits of the satellite, and (ii) more on retrograde orbits at high energies for initial retrograde orbits of the satellite. In any case, the most metal-rich stars are those of the thin disc, lost early and at high values of energy. 
These two features can be observed in the total distribution, which shows different regions dominated by metal-poor stars, depending on the initial satellite orbital properties.%\\

In the bottom panel of Figure \ref{fig:elzphi}, we show the same distributions in the case of an additional initial radial metallicity gradient. The in situ distributions are all very similar, this time showing the radial metallicity gradient in the region of the circular prograde orbits (thin disc). As far as the satellite distributions are concerned, unlike before, the most metal-poor stars are the ones of the thin disc, lost at low values of time and high values of energy. The distributions of satellite stars still show a dependence on the orbital parameter, $\phi_{orb}$, with more metal-poor stars on prograde orbits (negative values of $L_z$) at high energies for initial prograde orbits of the satellite and more on retrograde orbits at high energies for initial retrograde orbits of the satellite. Again, this can be appreciated in the total distribution, which shows different regions dominated by metal-poor stars, depending on the initial satellite orbital properties. %\\

This dependence could be useful to characterise the orbital parameters of the accretion events experienced by the MW. However, this can be challenging if the analysis is restricted to a volume of a few kiloparsecs in radius around the Sun (as is the case with many spectroscopic surveys). When we restrict the analysis to the solar-like volume, indeed, the final distributions in the $E-L_z$ space all become very similar to one another, losing the dependence of the metallicity pattern on the $\phi_{orb}$ parameter, since the differences in the final distribution considering the whole volume are due to satellite stars deposited at high energy values, which do not reach the solar vicinity. \\

%\newpage
As was mentioned above, the mean metallicity throughout the $E-L_z$ space for bound regions does not depend strongly on the orbital parameters of the accretion event. More quantitatively, we show in the upper row of Figure \ref{fig:MDFmeanphiEall} the mean metallicity (and standard deviation) in the energy interval $-5<E<0$ as a function of the angular momentum for all the possible initial values of the inclination of the satellite orbital plane (different colours) and for both the nograd case (left panels) and the grad case (right). 

We can see a dependence on the initial conditions assumed for the metallicity distribution of the satellite. In fact, both the left and right panels show a knee at approximately null angular momentum and then a plateau at low metallicity at more retrograde $L_z$ values (more positive). However, the nograd case also features a plateau at high metallicity and at prograde angular momentum values (negative $L_z$), whereas the grad case shows a decreasing mean metallicity value the more prograde the orbit. In the lower row of Figure \ref{fig:MDFmeanphiEall}, we also show the same plots in the energy interval $-10<E<-5$ for a comparison.

%\newpage
\subsection{Metallicity distribution functions in different regions of the $E-L_z$ space} 
\label{sec:mdf}

After discussing the variation in the mean metallicity across the $E-L_z$ space, the next step was to analyse the variation in the MDFs. The interest in looking at the MDF is due to the fact that all of the previous plots show the average metallicity in a given region of the $E-L_z$ plane, while the MDF can provide more insight into the population(s) which contribute(s) to a given average value. 
As before, to start off, we show the results for the simulation with an initial inclination of the satellite orbital plane with respect to the MW-type galaxy disc, $\phi_{orb}$, equal to $150^{\circ}$. The results concerning the simulations with direct, polar, and retrograde orbits ($\phi_{orb}=0^{\circ}, 90^{\circ}, 180^{\circ}$) are reported in the Appendix \ref{sec:elzmdfphi}.

\begin{table*}[h]
    \caption{Fraction of satellite stars with respect to the total in every cell of Figure \ref{fig:elzgrid}.}
    \label{tab:frac}    
    \resizebox{\textwidth}{!}{
    \centering
    \begin{tabular}{c|c c c c c c c c c c c }
         %\hline
         %\backslashbox{E}{$L_z$}
         $\downarrow E;  L_z \rightarrow$ & $(-110,-90)$ & $(-90,-70)$ & $(-70,-50)$ &$(-50,-30)$ &$(-30,-10)$ &$(-10,10)$ &$(10,30)$ &$(30,50)$ &$(50,70)$ &$(70,90)$ &$(90,110)$\\ \hline
         (0,5)      & - & - & - & 0.99 & 1.00 & 1.00 & 1.00 & 1.00 & 1.00 & 1.00 & 1.00\\
         (-5,0)     & - & - & - & - & 0.21 & 0.99 & 1.00 & 1.00 & 1.00 & 0.99 & -\\
         (-10,-5)   & - & - & - & - & 0.00 & 0.53 & 0.96 & 0.99 & - & - & -\\
         (-15,-10)  & - & - & - & - & - & 0.08 & 0.8 & - & - & - & -\\
         (-20,-15)  & - & - & - & - & - & 0.02 & - & - & - & - & -\\
         (-25,-20)  & - & - & - & - & - & - & - & - & - & - & -\\
         %\hline
    \end{tabular}}    
\end{table*}

\begin{figure}%[H]
    \centering
    \includegraphics[width=1\linewidth]{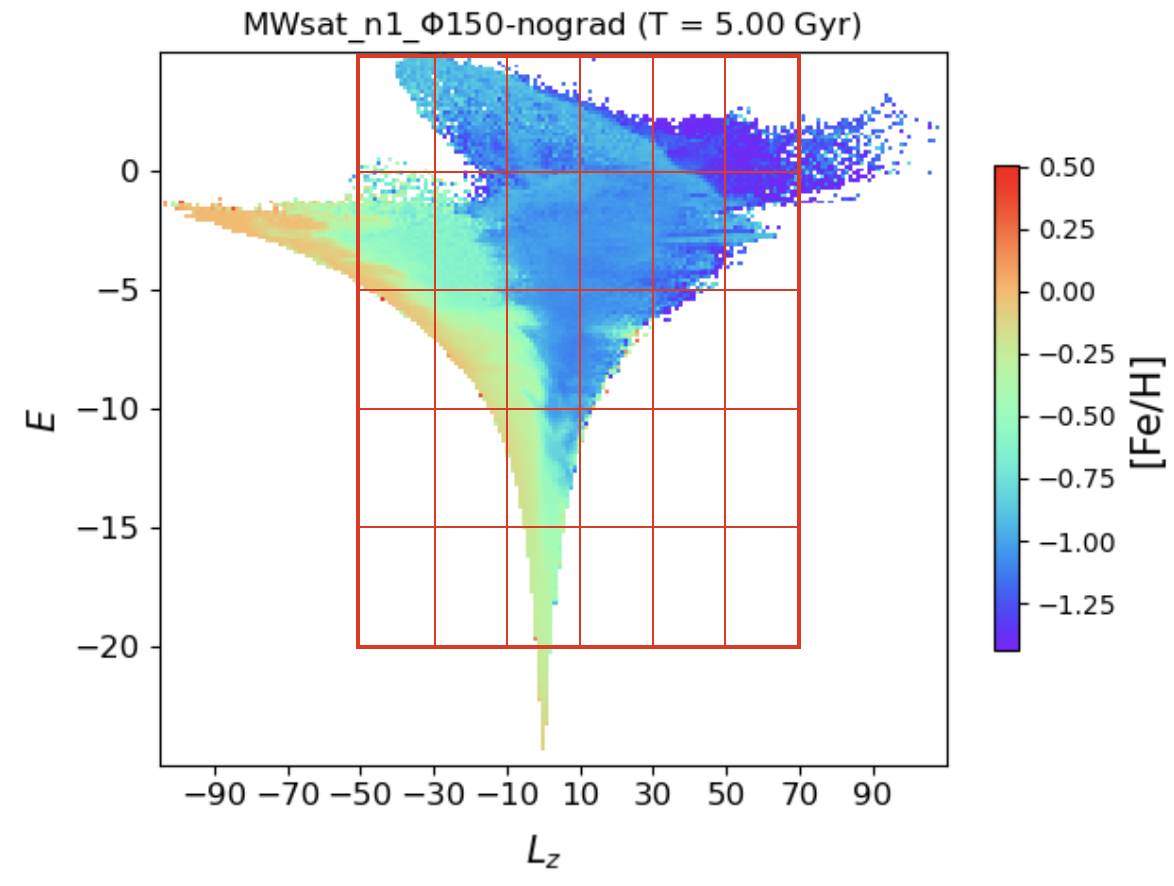}
    \includegraphics[width=1\linewidth]{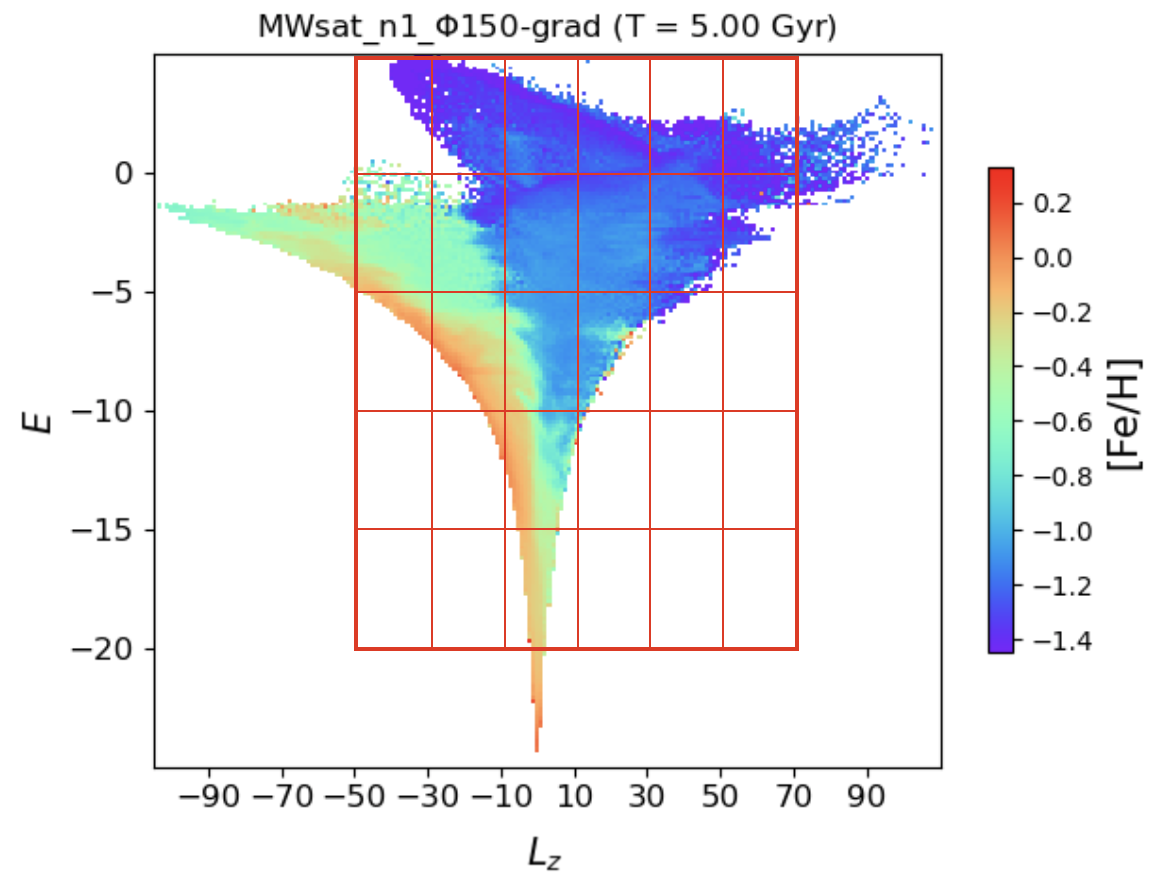}
    \caption{Total star distribution in the $E-L_z$ space in the case of an initial vertical metallicity gradient only (upper panel) and an additional initial radial metallicity gradient (lower panel), colour-coded at each point with the mean metallicity. The superimposed grid shows the splitting of the space used to analyse the MDFs in different regions.}
    \label{fig:elzgrid}
\end{figure}

\begin{figure*}%[H]
    \centering
    \includegraphics[width=1\linewidth]{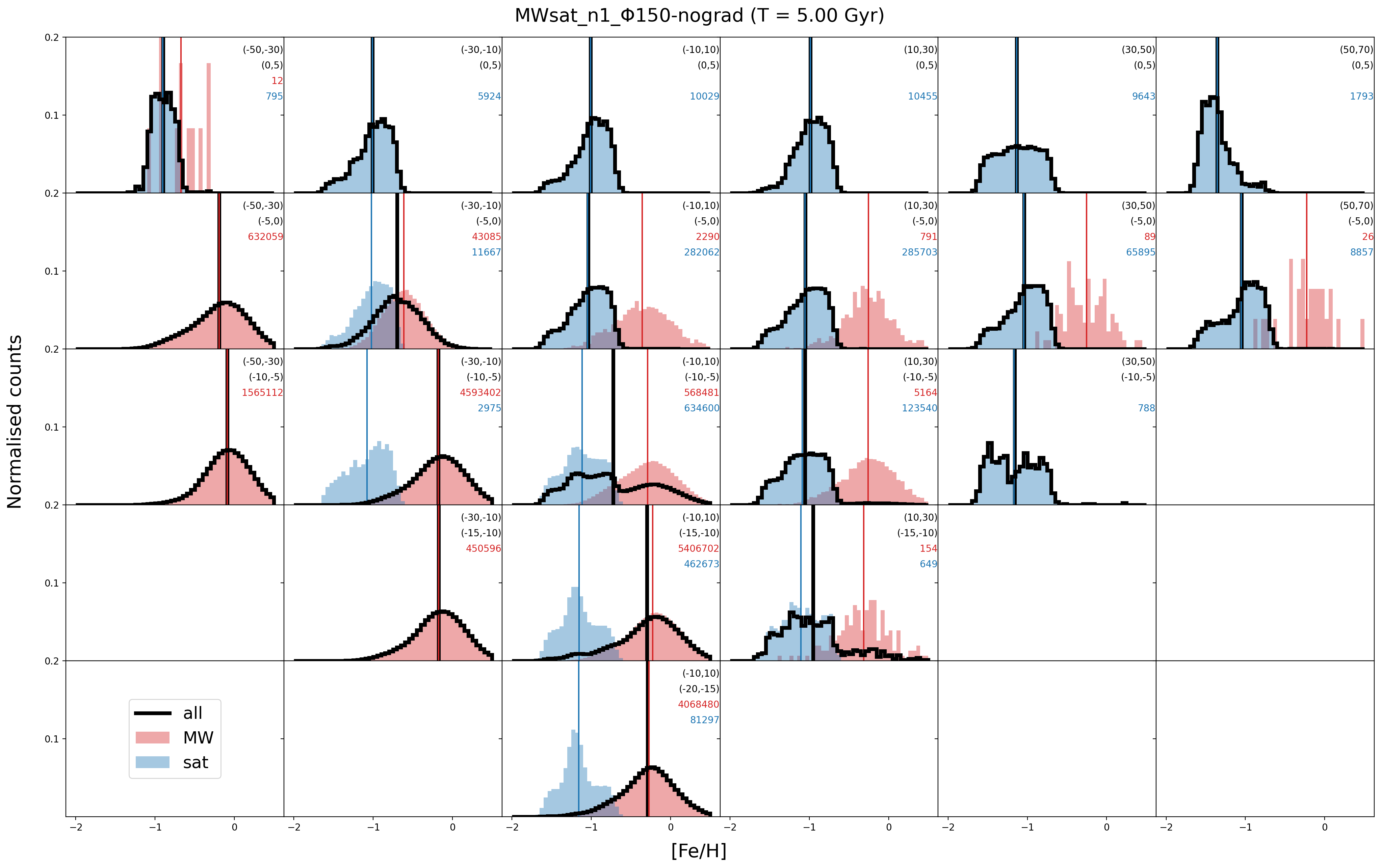}
    \includegraphics[width=1\linewidth]{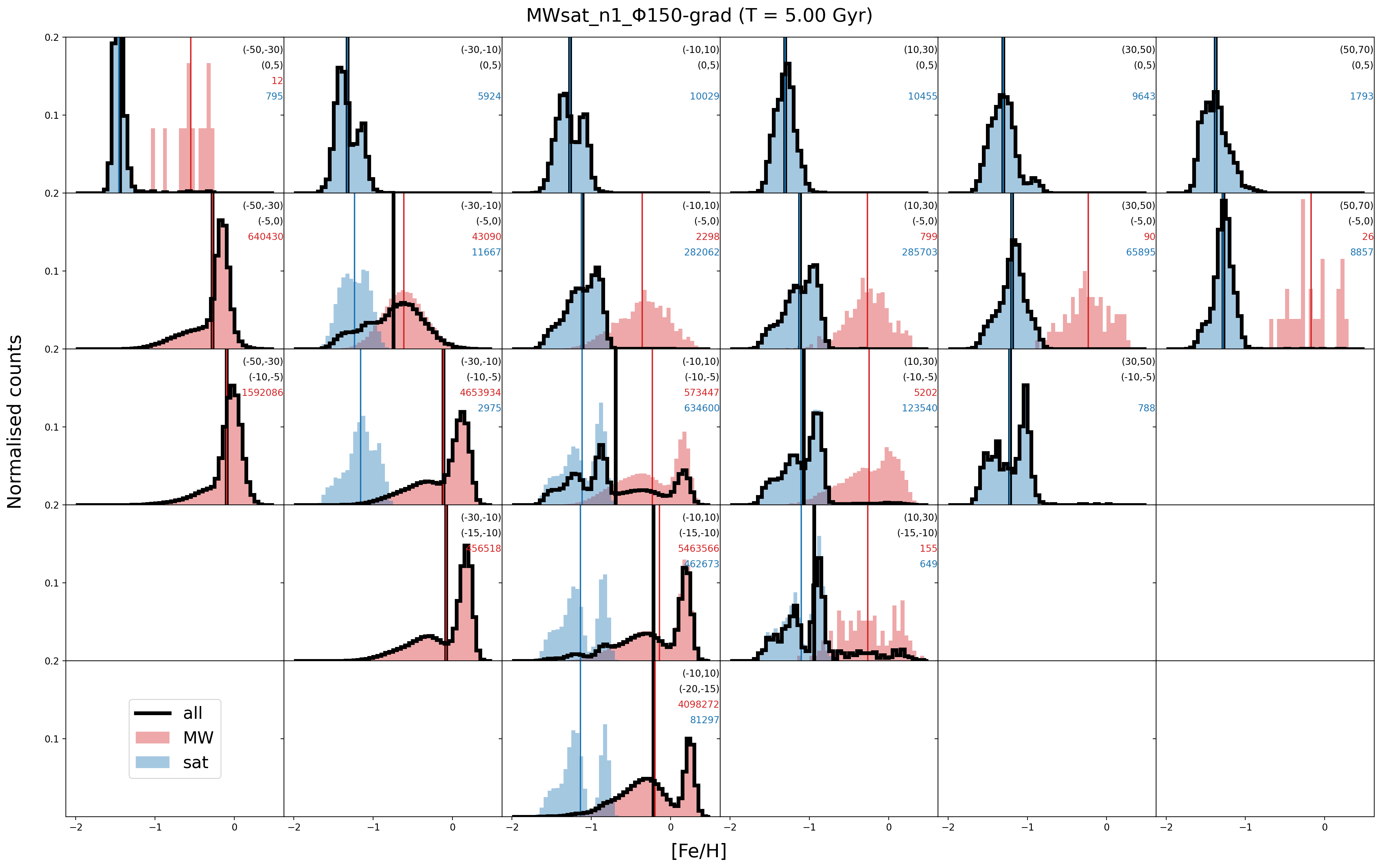}
    \caption{
    MDFs of different regions of the $E-L_z$, in the nograd case (top panel) and in the grad one (bottom panel). The red, blue, and black histograms show the MW-type galaxy, satellite, and total MDF, respectively, all normalised. The solid lines show the mean of the distribution. Energy intervals row by row (top to bottom): i) 0 < E < 5, ii) -5 < E < 0, iii) -10 < E < -5, iv) -15 < E < -10, v) -20 < E < -15. 
    Angular momentum intervals column by column (left to right): 
    i) -50 < $L_z$ < -30, ii) -30 < $L_z$ < -10, iii) -10 < $L_z$ < 10, iv) 10 < $L_z$ < 30, v) 30 < $L_z$ < 50, vi) 50 < $L_z$ < 70.
    The red (blue) number shows the amount of in situ (accreted) stars in every cell.
    }
    \label{fig:elzmdf}
\end{figure*}

In order to analyse the MDFs in different regions of the $E-L_z$ space, we divided this space into rectangles $20\times10^2$ kpc km/s wide in angular momentum and $5\times10^4$ km$^{2}$ s$^{-2}$ high in energy, so as to reproduce the average dimensions of the substructures found so far in the $E-L_z$ space. %We then obtained eleven bins in the \textit{x} axis and six bins in the \textit{y} axis
The splitting is shown in Figure \ref{fig:elzgrid}. 
In Figure \ref{fig:elzmdf}, we show the resulting MDFs for different regions of the $E-L_z$ space contained in the large red rectangle of Figure \ref{fig:elzgrid}. In this way, different rows show the MDFs for a fixed value of energy (increasing in the bottom-up direction), while different columns show the MDFs for a fixed value of angular momentum (increasing in the left-right direction). The angular momentum and energy intervals are reported in every cell, together with the number of in situ and accreted stars, in red and blue, respectively.
Every plot then shows the MDF considering all the stars in the region (in black), the MDF of the MW-type stars only (in red), and the MDF of the satellite stars only (in blue), all normalised in order to compare means, peaks, and shapes. The solid lines show the mean of the distribution, while the dotted ones show its peak(s).
The values of the fraction of satellite stars in the different $E-L_z$ regions, as well as the mean values of the total MDFs, in the two cases (nograd and grad) explored in this paper, are reported in Tables~\ref{tab:frac}, \ref{tab:mdfnograd} and \ref{tab:mdfgrad}, respectively.

\begin{figure*}[!htb]

    \includegraphics[width=1\linewidth]{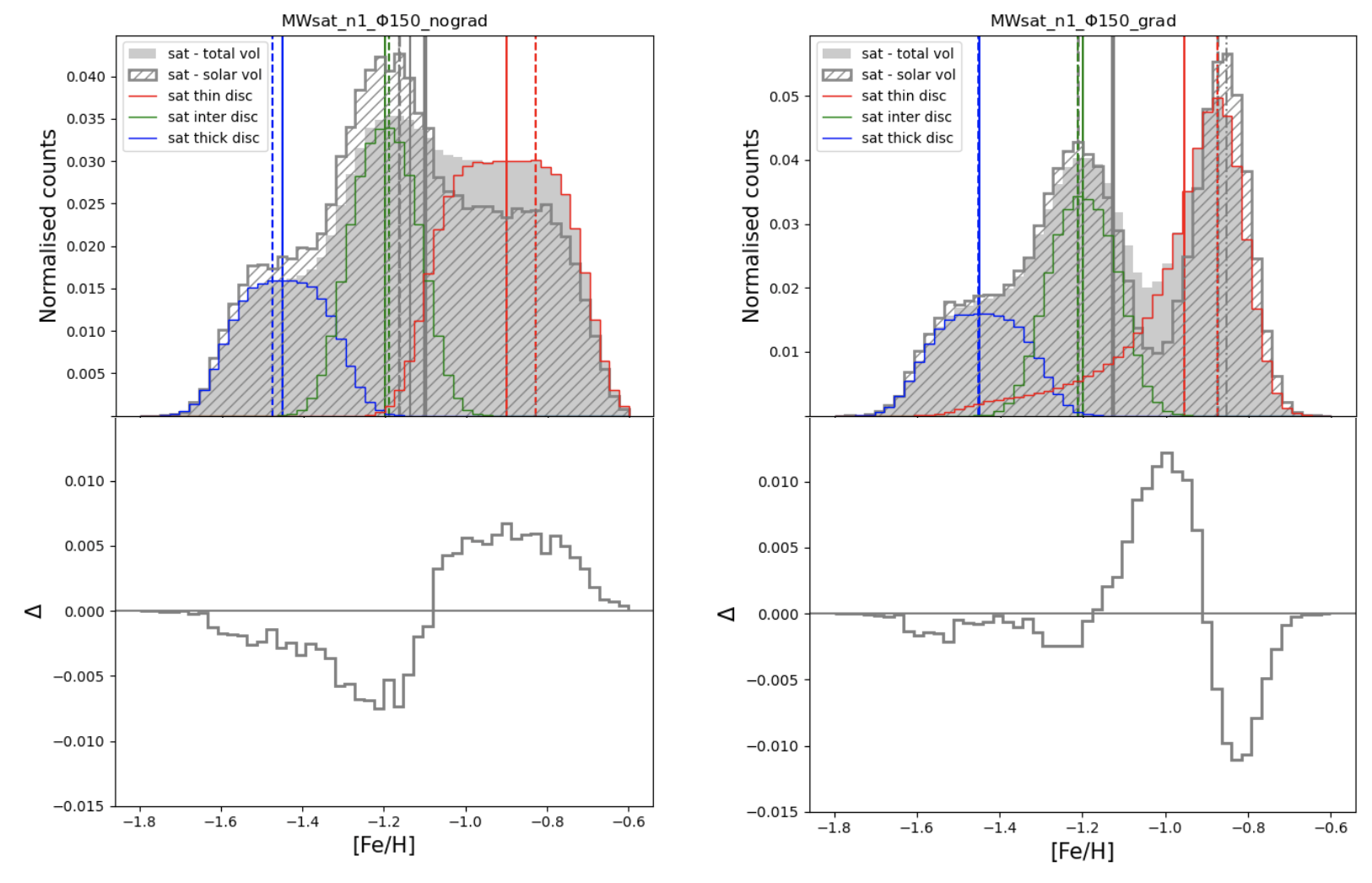} 
    \caption{Top panels: MDF of the satellite stars (in filled blue) and the same MDF restricted to the satellite stars that reach a solar-like volume by the end of the simulation (in patterned blue). The solid lines are the means of the distributions (broader for the total distribution, thinner for the solar one). The dashed (dash-dotted) lines are the peak(s) of the total (solar) distribution. The satellite MDF divided into disc components is shown by the red, green, and blue curves for the thin, intermediate, and thick discs, respectively. Bottom panel: Difference between the two normalised distributions in the top panel ($\Delta$), i.e. the total distribution minus the solar-like restricted one. The left column shows the case with the vertical metallicity gradient only, and the right column the one with the additional radial metallicity gradient.}
    \label{fig:sun}
\end{figure*}

We start with the case of the vertical metallicity gradient only. The $E-L_z$ space has been divided, as was described above and as is shown in the top panel of Figure \ref{fig:elzgrid}.
Starting from the bottom row of Figure \ref{fig:elzmdf}, the lowest energy interval considered ($-10 \le L_z \le 10, -20 \le E \le -15$), the global MDF in black of both accreted and in situ stars almost coincides with that of in situ stars only, since they dominate this region (see Figure \ref{fig:chem}).
The central rectangle of the following row ($-15 \le E \le -10$) is still dominated by in situ stars, but now the total MDF shows a bump at low metallicity ($\rm{[Fe/H]}\sim -1.2$) due to satellite stars. So far, satellite MDFs are all very similar one to each other, resembling the total satellite MDF (see Figure \ref{fig:chem}): there is a single-peaked distribution (at $\rm{[Fe/H]}\sim -1.2$), with a bump at higher metallicities.
The following row ($-10 \le E \le -5$) instead shows some differences. The satellite MDF starts to peak at higher metallicities (at $\rm{[Fe/H]}\sim -1$) in the central region with $L_z\sim0$, due to the fact that a larger fraction of satellite metal-rich stars are found at higher energies as they escape the satellite at earlier times. Moreover, we observe a broader distribution in the region with $-10 \le L_z \le 10$ and a region in which the MDF is double-peaked ($10 \le L_z \le 30$), even if the initial total MDF of the satellite was not.
We also see that the total MDF in regions with stars on more retrograde orbits has significant contributions from the satellite stars at low metallicities. As a consequence, its mean is shifted towards lower metallicities while moving towards more retrograde orbits (more positive values of $L_z$), but also towards higher energies, where satellite stars represent a higher fraction of the total (see Table \ref{tab:frac}). We can observe all these features in the least bound row ($-5 \le E \le 0$), where the distributions are broader and peaked at lower metallicities as we move towards more retrograde star orbits (positive $L_z$). In particular, the right-most plot of the figure, which contains the stars with the highest energies and most retrograde orbits, is the one where the MDF peaks at the lowest metallicities; that is, at $\rm{[Fe/H]}\sim -1.5$. For completeness, we also report the unbound energy interval $0<E<5$, even if these stars end up several hundreds kiloparsecs away from the main galaxy centre. We can also appreciate in this region the same trend; namely, the mean of the total MDF decreases for more retrograde orbits (see Tables \ref{tab:mdfnograd} and \ref{tab:mdfgrad}).

We now discuss the results in the case of an additional initial radial metallicity gradient, both in the satellite and in the MW-type galaxy.
We recall that the introduction of the radial gradient makes the MDFs of both the MW-type galaxy and the satellite double-peaked (see Figure \ref{fig:chem}). The highest peak is due to the thin disc component and the second  to the intermediate and thick disc components together (see the top panel of Figure \ref{fig:sun}, where we show the MDFs of the various disc components in different colours.) The in situ MDF peaks are at $\rm{[Fe/H]}\sim 0.2$ and at $\rm{[Fe/H]}\sim -0.2$, while the satellite MDFs are at $\rm{[Fe/H]}\sim -0.9$ and $\rm{[Fe/H]}\sim -1.2$. We were interested, then, in checking if we recover these features in the MDF restricted to different regions of the $E-L_z$ space. We proceeded in the same way as the previous case, so the $E-L_z$ space was divided as before and as is shown in the lower panel of Figure \ref{fig:elzgrid}.
In the bottom row of Figure \ref{fig:elzmdf}, we show the MDFs in the $E-L_z$ space with $E \le -15$. We can see that the MDF of the single galaxies, the MW-type galaxy (in red) and the satellite (in blue), are usually quite representative of the total initial ones, being bimodal and peaked at the initial positions in metallicities. Nevertheless, considering higher energy intervals ($-15 \le E \le -10$ and $-10 \le E \le -5$), the total MDF (in black) already starts to change, reflecting the increasing relative importance of the satellite contribution, and thus beginning to show bumps at lower metallicities and means more shifted towards lower metallicities too. This trend is hence independent of the choice of initial conditions for the metallicity gradient, but is mostly due to the fact that regions of higher energies and more retrograde angular momentum are more dominated by satellite stars (see Table \ref{tab:frac}), which are on average more metal-poor. 
In the regions with the highest energies (bound and not, $-5 \le E \le 0$ and $0 \le E \le 5$), we can see that the satellite MDFs change a lot in shape and mean, becoming broader, losing bimodality, and being on average at lower metallicities. Once again, the total MDF always coincides more with the satellite one, moving towards higher energies and more retrograde orbits, and thus changing a lot in mean towards lower metallicities.

We can conclude that, independently of the initial choice of metallicity gradient for the satellite, its MDFs in different regions of the $E-L_z$ space can be significantly different one from each other in shape, mean, and peak values, even if they all result from the same accretion event. Therefore, different regions of the $E-L_z$ space showing different kinematic and MDFs can actually be associated with the same single original accreted satellite, making the $E-L_z$ very hard to interpret. 
Nevertheless, we have found trends in the MDFs' varying energy and angular momentum values. In particular, we have shown that the mean and the peak(s) of the global MDF of both accreted and in situ stars move towards lower metallicities for higher values of energy and angular momentum (i.e. for more retrograde orbits). This is a consequence of the increasing importance of the more metal-poor satellite contribution in stellar density.

%\newpage

\begin{table*}%[h]
    \caption{Mean value of the metallicity distribution of all the stars for every cell of the top panel of Figure \ref{fig:elzgrid} (nograd case).} 
    \label{tab:mdfnograd}
    \centering
    \resizebox{0.8\textwidth}{!}{
    \begin{tabular}{c|c c c c c c c c }
         %\hline
         %\backslashbox{E}{$L_z$}
         $\downarrow E;  L_z \rightarrow$ %& $(-110,-90)$ & $(-90,-70)$ & $(-70,-50)$ 
         &$(-50,-30)$ &$(-30,-10)$ &$(-10,10)$ &$(10,30)$ &$(30,50)$ &$(50,70)$\\
         %&$(70,90)$ &$(90,110)$\\ 
         \hline
         (0,5)      & \makecell{-0.90} %\\MW: -0.68\\SAT: -0.905\\peak: -0.925 \\-0.825} 
                    & \makecell{-1.01} %\\MW: -\\SAT: -1.010\\peak: -0.925 \\-0.825}  
                    & \makecell{-1.01} %\\MW: -\\SAT: -1.006\\peak: -0.925\\-}  
                    & \makecell{-0.99} %\\MW: -\\SAT: -0.990\\peak: -1.025\\-}  
                    & \makecell{-1.13} %\\MW: -\\SAT: -1.132\\peak: -1.275 \\-0.875}  
                    & \makecell{-1.36}\\ %\\MW: -\\SAT: -1.357\\peak: -1.525 \\-0.825} \\
         %\hline 
         (-5,0)     & \makecell{-0.19} %\\MW: -0.190\\SAT: -\\peak: -0.075}  
                    & \makecell{-0.69} %\\MW: -0.611\\SAT: -1.020\\peak: -0.725}  
                    & \makecell{-1.04} %\\MW: -0.340\\SAT: -1.045\\peak: -0.875}  
                    & \makecell{-1.05} %\\MW: -0.230\\SAT: -1.053\\peak: -0.875}  
                    & \makecell{-1.04} %\\MW: -0.204\\SAT: -1.043\\peak: -0.925}  
                    & \makecell{-1.05}\\ %\\MW: -0.116\\SAT: -1.050\\peak: -0.875} \\
         %\hline
         (-10,-5)   & \makecell{-0.09} %\\MW: -0.091\\SAT: -\\peak: -0.075\\-\\-}  
                    & \makecell{-0.18} %\\MW: -0.178\\SAT: -1.071\\peak: -0.125\\-\\-}  
                    & \makecell{-0.72} %\\MW: -0.288\\SAT: -1.114\\peak: -1.175 \\-0.825 \\-0.225}  
                    & \makecell{-1.06} %\\MW: -0.257\\SAT: -1.088\\peak: -0.875\\-\\-}  
                    & \makecell{-1.16} %\\MW: -\\SAT: -1.168\\peak: -1.375 \\-1.025 \\-0.925}  
                    & \makecell{-}\\ %\\MW: -\\SAT: -\\peak: -\\-} \\
         %\hline 
         (-15,-10)  & \makecell{-} %\\MW: -\\SAT: -\\peak: -\\-}  
                    & \makecell{-0.17} %\\MW: -0.172\\SAT: -\\peak: -0.125\\-}  
                    & \makecell{-0.29} %\\MW: -0.138\\SAT: -1.151\\peak: -0.175\\-}  
                    & \makecell{-0.96} %\\MW: -0.303\\SAT: -1.111\\peak: -1.275\\ 
                    & \makecell{-} %\\MW: -\\SAT: -\\peak: -\\-}  
                    & \makecell{-}\\ %\\MW: -\\SAT: -\\peak: -\\-} \\
         %\hline
         (-20,-15)  & \makecell{-} %\\MW: -\\SAT: -\\peak: -}  
                    & \makecell{-} %\\MW: -\\SAT: -\\peak: -}  
                    & \makecell{-0.29} %\\MW: -0.272\\SAT: -1.159\\peak: -0.225}  
                    & \makecell{-} %\\MW: -\\SAT: -\\peak: -}  
                    & \makecell{-} %\\MW: -\\SAT: -\\peak: -}  
                    & \makecell{-}\\ %\\MW: -\\SAT: -\\peak: -} \\
         %\hline
    \end{tabular}
    }    
\end{table*}

\begin{table*}%[h]
    \caption{Mean value of the metallicity distribution of all the stars for every cell of the bottom panel of Figure \ref{fig:elzgrid} (grad case).}
    \label{tab:mdfgrad}
    \centering
    \resizebox{0.8\textwidth}{!}{
    \begin{tabular}{c|c c c c c c c c }
         %\hline
         %\backslashbox{E}{$L_z$}
         $\downarrow E;  L_z \rightarrow$ %& $(-110,-90)$ & $(-90,-70)$ & $(-70,-50)$ 
         &$(-50,-30)$ &$(-30,-10)$ &$(-10,10)$ &$(10,30)$ &$(30,50)$ &$(50,70)$\\
         %&$(70,90)$ &$(90,110)$\\ 
         \hline
         (0,5)      & \makecell{-1.45} %\\MW: -0.653\\SAT: -1.459\\peak: -1.475\\-} 
                    & \makecell{-1.33} %\\MW: -\\SAT: -1.328\\peak: -1.375 \\    -1.125}  
                    & \makecell{-1.27} %\\MW: -\\SAT: -1.270\\peak: -1.325 \\    -1.125}  
                    & \makecell{-1.31} %\\MW: -\\SAT: -1.306\\peak: -1.275\\ -}  
                    & \makecell{-1.31} %\\MW: -\\SAT: -1.306\\peak: -1.275 \\    -0.925}  
                    & \makecell{-1.38} \\ %\\MW: -\\SAT: -1.375\\peak: -1.475 \\    -1.375} \\
         %\hline 
         (-5,0)     & \makecell{-0.28} %\\MW: -0.277\\SAT: -\\peak: -0.175\\-}  
                    & \makecell{-0.74} %\\MW: -0.610\\SAT: -1.234\\peak: -0.575\\-}  
                    & \makecell{-1.11} %\\MW: -0.358\\SAT: -1.118\\peak: -0.925\\-}  
                    & \makecell{-1.12} %\\MW: -0.249\\SAT: -1.121\\peak: -1.175 \\-0.925}  
                    & \makecell{-1.19} %\\MW: -0.261\\SAT: -1.193\\peak: -1.175\\-}  
                    & \makecell{-1.28}\\ %\\MW: -0.114\\SAT: -1.284\\peak: -1.275\\-} \\
         %\hline
         (-10,-5)   & \makecell{-0.09} %\\MW: -0.099\\SAT: -\\peak: -0.025\\-\\-}  
                    & \makecell{-0.12} %\\MW: -0.116\\SAT: -1.156\\peak: 0.125\\-\\-}  
                    & \makecell{-0.69} %\\MW: -0.230\\SAT: -1.113\\peak: -1.225 \\-0.875  \\0.175}  
                    & \makecell{-1.07} %MW: -0.242\\SAT: -1.105\\peak: -1.175 \\-0.875\\-}  
                    & \makecell{-1.22} %\\MW: -\\SAT: -1.227\\peak: -1.475 \\-1.025\\-}  
                    & \makecell{-} \\%\\MW: -\\SAT: -\\peak: -\\-\\-} \\
         %\hline 
         (-15,-10)  & \makecell{-} %\\MW: -\\SAT: -\\peak: -\\-\\-}  
                    & \makecell{-0.08} %\\MW: -0.080\\SAT: -\\peak: -0.325 \\0.175\\-}  
                    & \makecell{-0.22} %\\MW: -0.138\\SAT: -1.134\\peak: -0.325  \\0.175\\-}  
                    & \makecell{-0.95} %\\MW: -0.303\\SAT: -1.103\\peak: -1.475 \\-1.225 \\-0.875 %+-0.425 +-0.275  +0.075
                    & \makecell{-} %\\MW: -\\SAT: -\\peak: -\\-\\-}  
                    & \makecell{-} \\%\\MW: -\\SAT: -\\peak: -\\-\\-} \\
         %\hline
         (-20,-15)  & \makecell{-} %\\MW: -\\SAT: -\\peak: -\\-}  
                    & \makecell{-} %\\MW: -\\SAT: -\\peak: -\\-}  
                    & \makecell{-0.21} %\\MW: -0.196\\SAT: -1.135\\peak: -0.325 \\0.225 }  
                    & \makecell{-} %\\MW: -\\SAT: -\\peak: -\\-}  
                    & \makecell{-} %\\MW: -\\SAT: -\\peak: -\\-}  
                    & \makecell{-} \\%\\MW: -\\SAT: -\\peak: -\\-} \\
         %\hline
    \end{tabular}
    }    
\end{table*}

\paragraph{Global versus local metallicity distribution functions} 
%\subsubsection{Global versus local metallicity distribution functions} 
\label{sec:elzmdfsun}

So far, we have shown MDFs estimated over the full galactic extent, in the ideal case -- as in a simulation -- in which all stars (or stellar particles) in the galaxy can be observed. In reality, spectroscopic surveys provide chemical abundances over a limited volume around the Sun. It is thus interesting to understand to what extent ‘local’ MDF can be representative of the global one; in particular, if we were able to select a pure accreted sample of stars (that is, with null or very limited pollution of in situ stars) a few kiloparsecs from the Sun, the extent to which its MDF would be representative of that of the whole satellite. \\

We recall that we defined the ‘solar-vicinity’ as a spherical volume centred at 8 kpc from the MW-type galaxy centre and with a radius of 5 kpc.
The top panels of Figure \ref{fig:sun} show the MDF of all satellite stars (in filled gray) and its local counterpart, in which only the contribution of stars that at the end of the simulation are found in a solar-like volume are shown (in patterned gray). Both the cases with and without an initial radial metallicity gradient in the satellite are shown (right and left panels, respectively). We can notice that in both cases the global shape of the MDF is conserved, as the metallicity range featured is also.

The bottom panels show $\Delta$, the difference between the two normalised distributions. We can see that in the nograd case there is an under-representation of high-metallicity stars (i.e. stars with $\rm{[Fe/H]} \ge -1.1$), while in the grad case there is a deeper drop at $\rm{[Fe/H]} \sim -1.0$. 

By way of a comparison with the MDFs of the single disc components (thick, intermediate, and thin ones in red, green, and blue, respectively, in the top panel), we can associate the under-representation with missing thin disc stars; that is, with stars initially on the outskirts of the satellite that are lost in the early phases of the accretion, and that are deposited at energies too high to be sampled in the solar vicinity. Depending on the metallicity that these stars initially have, this will give rise to an under-representation of these metallicities in local MDFs.

%\newpage
\subsection{Reconstructing a possible (but wrong) accretion history from kinematic spaces and metallicity distribution functions} 
\label{sec:kirby}

The results discussed in the previous section can lead one to reconstruct accretion histories that may look realistic, but that are in fact very far from true. An example of the difficulty of using these spaces to reconstruct the merger tree of a MW-type galaxy is given below.\\
Suppose we have a dataset of halo stars, distributed over an extended range of energies and angular momenta. We might be led to interpret different regions of this space as being dominated by stars of different origins, and to infer the properties of the galaxies in which these stars formed by studying the characteristics of their MDFs. For example, by making use of a mass-metallicity relation given by \citep{kirby2013},
\begin{equation}
    <\rm{[Fe/H]}> = (-1.69 \pm 0.04) + (0.30 \pm 0.02) \log\left(\frac{M_*}{10^6M_{\odot}}\right),
    \label{eq:kirby}
\end{equation}
we may in principle reconstruct the stellar mass of the progenitor galaxies, given the mean metallicity of stars in a given $E-L_z$ region (Tables \ref{tab:mdfnograd} and \ref{tab:mdfgrad}).
If we apply this reasoning to our simulations, we would get the following:
\begin{itemize}
\item Unbound stars ($0<E<5$) found in the nograd simulation  (upper panel of Figure \ref{fig:elzgrid}) are dominated at null angular momentum ($-10<L_z<10$) by stars of a galaxy whose stellar mass was $\sim 10^{8.4}  M_{\odot}$, since the mean metallicity of stars in this region is $<\rm{[Fe/H]}>\sim-1.0$, while stars of the same energy but more retrograde angular momenta ($50<L_z<70$) would be associated with a satellite galaxy with a stellar mass equal to $10^7 M_{\odot}$, given that the mean stellar metallicity in this region is $<\rm{[Fe/H]}>\sim-1.4$.
\item At lower energies ($-10<E<-5$), the problem would persist. In particular, in the region around null angular momentum ($-10<L_z<10$), the mean metallicity being $<\rm{[Fe/H]}>\sim -0.7$, this may be interpreted as a signature of a massive accretion, $M_* \sim 10^{9.4}  M_{\odot}$.
\item Similar conclusions would hold for the grad case, in which, depending on the region we look at, we may end up finding traces of low-mass accretions ($M_* \sim 10^{7.4}  M_{\odot}$ at $-5<E< 0$ and $50<L_z<70$, corresponding to a mean of $<\rm{[Fe/H]}>\sim -1.3$) and also more massive ones  ($M_* \sim 10^{8.4}  M_{\odot}$ at $-10<E<-5$ and $10<L_z<30$, corresponding to a mean of $<\rm{[Fe/H]}>\sim -1.0$).
\end{itemize}
An even more complex situation would arise if we tried to interpret the multiple peaks present in the MDFs of some specific regions of the $E-L_z$ space as being due to distinct accretion events.\\

Regardless of the details of the average metallicity values, and thus the masses of the progenitors, the common feature of this type of analysis is to overestimate the number of accretions and underestimate the associated masses.

%\newpage
\subsection{Metallicity gradient in the satellite versus in the $E-L_z$ space}
\label{sec:relgrad}

Once it had been established that analysing separately the metallicity distribution of limited regions inside the kinematic spaces is not necessarily informative about the initial conditions of the original satellite, we tried to consider the global picture in order to characterise the accretion event. To this end, we set the initial conditions for both the galaxies to the grad case, but while for the MW-type galaxy it was fixed as is described in Section \ref{sec:methods}, for the satellite we let the steepness of the radial metallicity gradient vary in the range: $f_{satthin} \in [-0.06, 0.02]$. In this way, we could evaluate the dependence on this parameter of the metallicity patterns in the $E-L_z$ space, without pre-selecting a certain group of accreted stars on the basis of kinematic arguments. 

\begin{figure}  

    \includegraphics[width=1.1\linewidth]{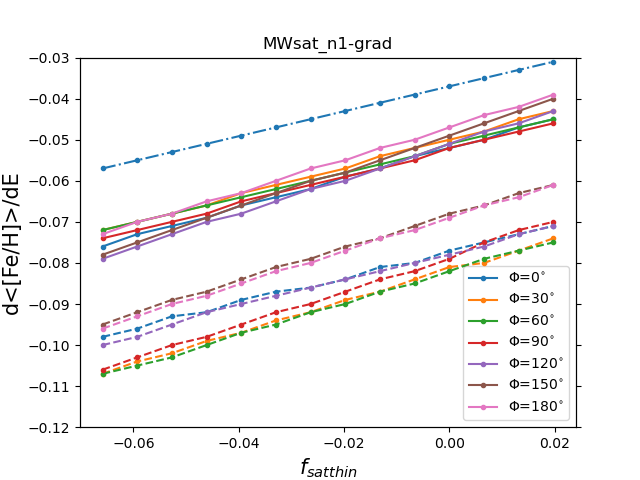} 
    \caption{Relation between the slope of the initial radial metallicity gradient in the satellite thin disc ($f_{satthin}$) and the slope of the metallicity gradient in the final distribution of all the thin disc stars in the $E-L_z$ space as a function of energy (d<[Fe/H]>/dE), in the total volume of the simulation (solid lines), and in the restricted solar-like volume (dashed line).
    The case of $\phi_{orb}=0^{\circ}$ also features the dash-dotted curve described in Section \ref{sec:relgrad}.}
    \label{fig:slope}
\end{figure}

For every $f_{satthin}$ value, we considered the final distribution of both satellite and MW-type thin disc stars in the $E-L_z$ space and we computed the mean metallicity as a function of the energy. We generally found a monotonic decreasing function as the energy increases, and thus we fit it with a line (more details about the method can be found in Appendix \ref{sec:fit}) and analysed the relation between the slope of this linear fitting ($d<\rm{[Fe/H]}>/dE$) and $f_{satthin}$ itself. The relation is shown by the solid lines in Figure \ref{fig:slope}, in different colours for the different orbital parameter values (see captions). 
We can see that it is a linear relation, which is independent of the initial inclination of the satellite orbital plane, and thus from the metallicity gradient in the kinematic space as a function of energy one can obtain information about the initial conditions of the metallicity gradient in the original satellite. \\

There is only one exception, which is the case of $\phi_{orb}=0^{\circ}$ and which is shown by the dash-dotted line in Figure \ref{fig:slope}. This is due to the fact that for this planar orbit satellite stars reach very bound orbits, and thus are redistributed at very low energy values in the $E-L_z$ space. As a consequence, they dump the mean metallicity at low energy, which then becomes an increasing function of energy in the lowest interval, $-25<E<-22$, and starts to decrease as before after this value. If we then fit the relation with a cut in energy to include only the decreasing range, we obtain the solid blue line that is consistent with all the other curves for the different inclinations. \\

As a final step, we performed the same analysis considering the final distribution in the $E-L_z$ space of the thin disc stars that reach the solar vicinity. The relations for every $\phi_{orb}$ value are shown by the dashed lines in Figure \ref{fig:slope}. We obtain the same relation, yet shifted to more negative values of the slope of the metallicity gradient in the final distribution in the $E-L_z$ space as a function of energy ($d<\rm{[Fe/H]}>/dE$). \\

We have shown that merger debris is not clustered, making it challenging to derive merger parameters due to variations in the MDF across different regions of $E-L_z$. This limitation may seem to render the IOM space less useful. However, there is a positive aspect to consider. Thanks to the energy differentiation of the debris, we gain insights into the structure of the galaxy-progenitors, as is demonstrated through the gradient relation in this analysis \citep[see also][]{khoperskov23d}. Moreover, this would not be achievable if mergers were viewed as $E-L_z$ clumps.

%\newpage 
%--------------------------------------------------------------------
\section{Conclusions}
%-----------------------------------------------------------------
\label{sec:conclusions}

In this paper, we have analysed N-body simulations of the accretion of a satellite galaxy onto a MW-type galaxy (an accretion with a mass ratio of 1:10), to explore the extent to which we can couple kinematic characteristics and metallicities of stars in the halo to reconstruct the accretion history of our Galaxy. The main conclusions of our work are the following:
\begin{itemize}
 
   \item We confirm earlier results  that accreted stars from $\sim$1:10 mass ratio satellites are redistributed in a wide range of energies and angular momenta \citep{jean-baptiste2017, koppelman20, amarante22, khoperskov23b} and that in particular stars deposited at higher energies have on average different metallicities than those of stars that at the end of the accretion process end up being more gravitationally bound to the MW \citep[][]{amarante22, khoperskov23c, khoperskov23d}.
\item Because satellite stars with different metallicities can be deposited in different regions of the $E-L_z$ space, this implies that a single accretion of $\sim$1:10 can manifest with different MDFs, in different regions of the $E-L_z$ space. 
\item  Groups of stars with different $E$, $L_z$, and metallicities may be interpreted as originating from different satellite galaxies, but our analysis shows that these interpretations are not physically motivated.
\item When the analysis is restricted to a solar-like volume, we can no longer capture the very metal-rich (or metal-poor, depending on the initial conditions in the satellite) accreted stars found at very high energies. 
This implies that the MDF of the GSE may still exhibit biases. 
\item Reconstructing the Galactic accretion history through mass-metallicity relations using the MDF mean values of single clumps can be very misleading. For example, a single merger of a galaxy with a mass of $5\times10^{10}$ can be erroneously interpreted as three with masses of $\sim10^{7}$, $\sim10^{8.4}$, and $\sim10^{9.4}$.
As we show, coupling kinematic information with the MDFs to reconstruct the accretion history of the MW can bias the reconstructed merger tree towards increasing the number of past accretions and decreasing the masses of the progenitor galaxies.
\item From the metallicity gradient in the $E-L_z$ space as a function of energy, one can retrieve information about the initial conditions of the radial metallicity gradient in the original satellite.
\end{itemize}

A number of possible accretion events have been proposed in the literature by making use of this approach, and our analysis reinforces previous suggestions that at least some of these accretion events are disputable. Indeed, we have shown how a single massive accretion can produce MDFs whose characteristics change across the $E-L_z$ plane. In a scenario in which the MW may have experienced more than a single significant accretion in its past, it is clear that the interpretation of these spaces, and related MDFs, becomes even more challenging. In this respect, we recall that even the GSE merger initially discovered in a solar vicinity volume \citep{nissenschuster2010} was already suspected to have been made by more than one galaxy at the time of its local discovery \citep[see discussion in][]{nissenschuster2010}. New evidence has recently been published on the possibility that this accretion event indeed hides multiple ones \citep{donlon23, nissen23}. 

Ultimately, overall the merger tree of the MW is still far from being robustly established, and - even more importantly - we still need to establish sound methods of discriminating between accretion scenarios. 
Although various abundance ratios have already been used to characterise different accretion events \citep[e.g.,][]{matsuno19, monty20, matsuno22a, matsuno22b, horta23, ceccarelli24}, our study emphasises the importance of going not only beyond [Fe/H] but also a revision of the kinematic identification of merger debris. This will be vital for a more realistic reconstruction of the MW merger history in light of coming data from WEAVE, 4MOST, SDSSV, and other spectroscopic surveys.

\begin{acknowledgements}
AM and SS acknowledge support from the ERC Starting Grant NEFERTITI H2020/808240. This work has made use of the computational resources obtained through the DARI grant A0120410154 (P.I.: P. Di Matteo). The authors thank P. Bonifacio for insightful discussion.
\end{acknowledgements}

%--------------------------------------------------------------------
% bibliography
%--------------------------------------------------------------------
\bibliographystyle{aa}
\bibliography{aa}

%--------------------------------------------------------------------
% appendix
%--------------------------------------------------------------------
    
%\begin{comment}

\begin{appendix}
\section{Distributions in $E-L_z$ space for different disc components}
\label{sec:discs}

\begin{figure*}%[H]
    \includegraphics[width=1\linewidth]{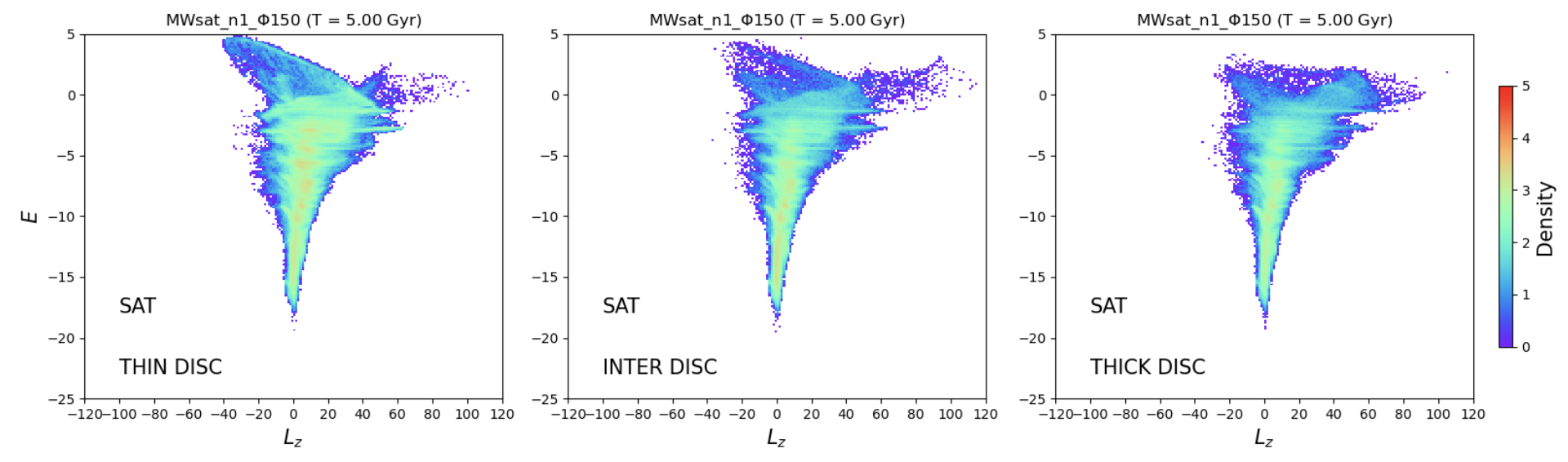}    
    \includegraphics[width=1\linewidth]{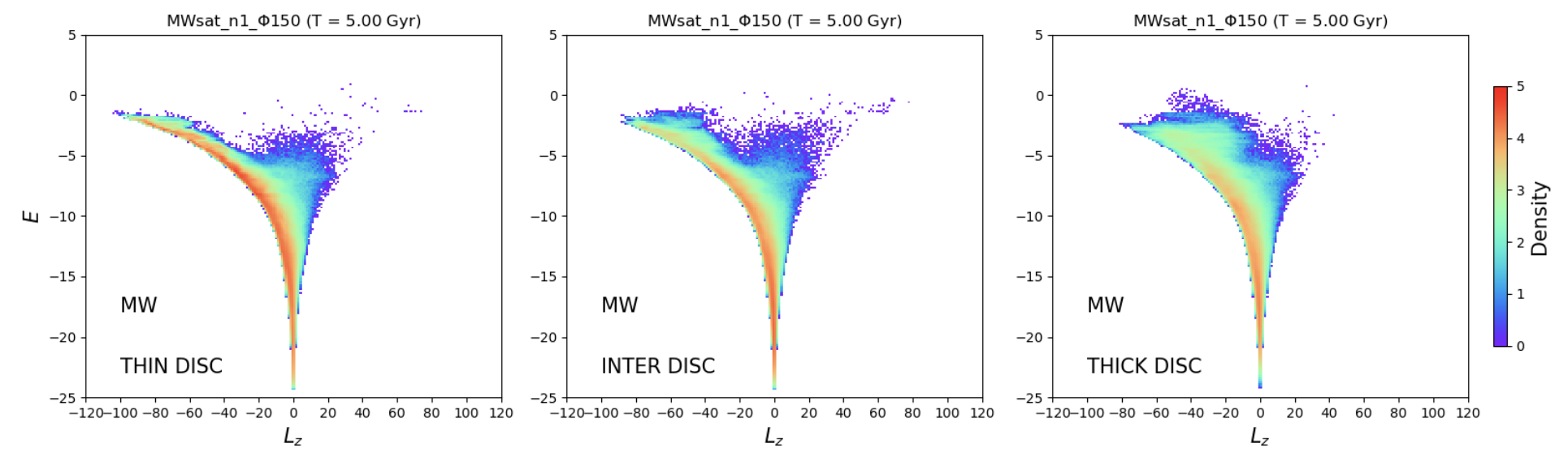}
    \caption{Final distributions in the $E-L_z$ space for different disc components (thin, intermediate, and disc components in the first, second, and third columns, respectively), for satellite stars in the first row and in situ stars in the second one.}
    \label{fig:discs}
\end{figure*}

In Figure \ref{fig:discs}, we report -- for the simulation with the initial inclination of the satellite orbital plane of $\phi_{orb}=150^{\circ}$ analysed in the paper -- the final distribution in the $E-L_z$ space of the satellite (top row) and in situ (bottom row) stars, divided into their disc components: thin, intermediate, and thick disc stars in the first, second, and third columns, respectively.

Since the different disc components have been assigned different metallicity values, these distributions are crucial to interpreting the metallicity patterns and gradients that we observe at the end in the $E-L_z$ space.

\section{ Metallicity distribution function in different regions of the $E-L_z$ space for different $\phi_{orb}$}
\label{sec:elzmdfphi}

\begin{figure*}[h]
    \includegraphics[width=1\linewidth]{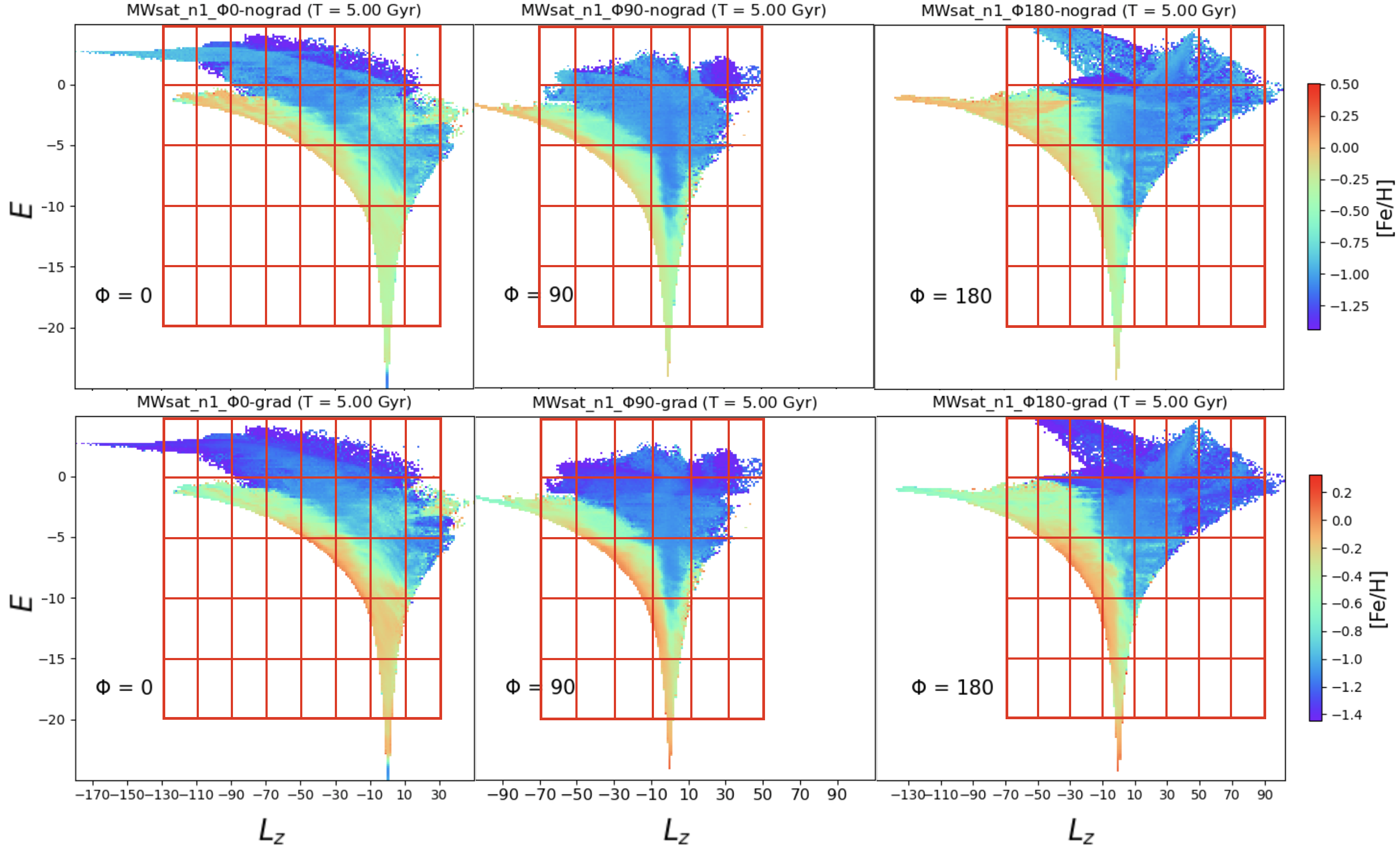}  
    \caption{Same as Figure \ref{fig:elzgrid}, but for the case of $\phi_{orb}=0^{\circ}, 90^{\circ}, 180^{\circ}$ in the first, second, and third columns, respectively. The first row shows the case of an initial vertical metallicity gradient only and the second row that with an additional initial radial metallicity gradient.}
    \label{fig:elzgrids}
\end{figure*}

We wanted to check the dependence of the results concerning the MDF analysis on the orbital parameter, $\phi_{orb}$; that is, the initial inclination of the satellite orbital plane. We considered three more cases for a planar, polar, and retrograde orbit, $\phi_{orb}=0^{\circ}, 90^{\circ}, 180^{\circ}$, and we performed the same analysis made in the $\phi_{orb}=150^{\circ}$ case (described in Section \ref{sec:mdf}). 

In Figure \ref{fig:elzgrids}, we show the final total distribution in the $E-L_z$ plane colour-coded by the mean metallicity at each point (described in Section \ref{sec:patterns}), in which the different columns correspond to $\phi_{orb}=0^{\circ}, 90^{\circ}, 180^{\circ}$ and the two different rows correspond to the nograd and grad cases, respectively. The grid shows the splitting of the $E-L_z$ regions in which we analysed the MDFs, and the red rectangles indicate the ones reported in the following Figures, \ref{fig:elzmdf0}, \ref{fig:elzmdf90}, and \ref{fig:elzmdf180}. 

Every panel of Figures, \ref{fig:elzmdf0}, \ref{fig:elzmdf90}, and \ref{fig:elzmdf180} then shows the MDF, considering all the stars in the region (in black), the MDF of the MW-type stars only (in red), and the MDF for the satellite stars only (in blue), all normalised in order to compare means, peaks, and shapes. The solid lines show the mean of the distribution.
In all these figures, different rows show the MDFs for a fixed interval of energy, which from top to bottom are: i) 0 < E < 5, ii) -5 < E < 0, iii) -10 < E < -5, iv) -15 < E < -10, and v) -20 < E < -15.%, vi) -25 < E < -20.
The different columns, on the other hand, show the MDFs for a fixed interval of angular momentum (increasing from left to right). 
The angular momentum intervals, column by column, are, from left to right: for Figure \ref{fig:elzmdf0}, i) -130 < $L_z$ < -110, ii) -110 < $L_z$ < -90, iii) -90 < $L_z$ < -70, iv) -70 < $L_z$ < -50, v) -50 < $L_z$ < -30, vi) -30 < $L_z$ < -10, vii) -10 < $L_z$ < 10, and viii) 10 < $L_z$ < 30; for Figure \ref{fig:elzmdf90}, i) -70 < $L_z$ < -50, ii) -50 < $L_z$ < -30, iii) -30 < $L_z$ < -10, iv) -10 < $L_z$ < 10, and v) 10 < $L_z$ < 30, vi) 30 < $L_z$ < 50; and for Figure \ref{fig:elzmdf180}, i) -70 < $L_z$ < -50, ii) -50 < $L_z$ < -30, iii) -30 < $L_z$ < -10, iv) -10 < $L_z$ < 10, v) 10 < $L_z$ < 30, vi) 30 < $L_z$ < 50, vii) 50 < $L_z$ < 70, and viii) 70 < $L_z$ < 90.

The aim is to analyse the MDF in different regions of the $E-L_z$ space, in order to find any trend in their means, peaks, and overall shape as a function of the region considered; for instance, fixing the value of the energy and varying the one of the angular momentum, and vice versa. 
All the simulations show a lower value of mean metallicity the higher the energy.

\begin{figure*}%[H]
    \includegraphics[width=1\linewidth]{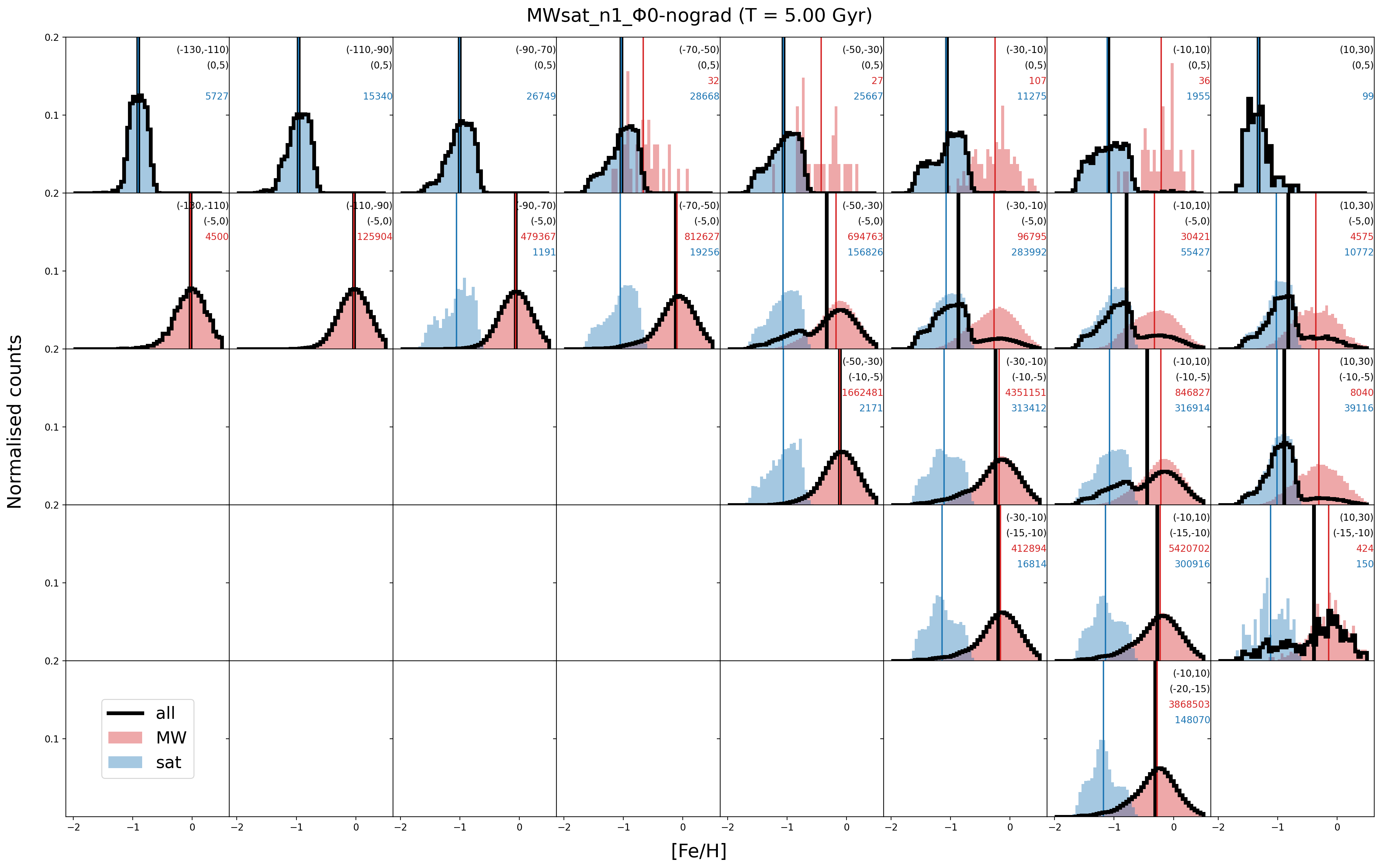}    
    \includegraphics[width=1\linewidth]{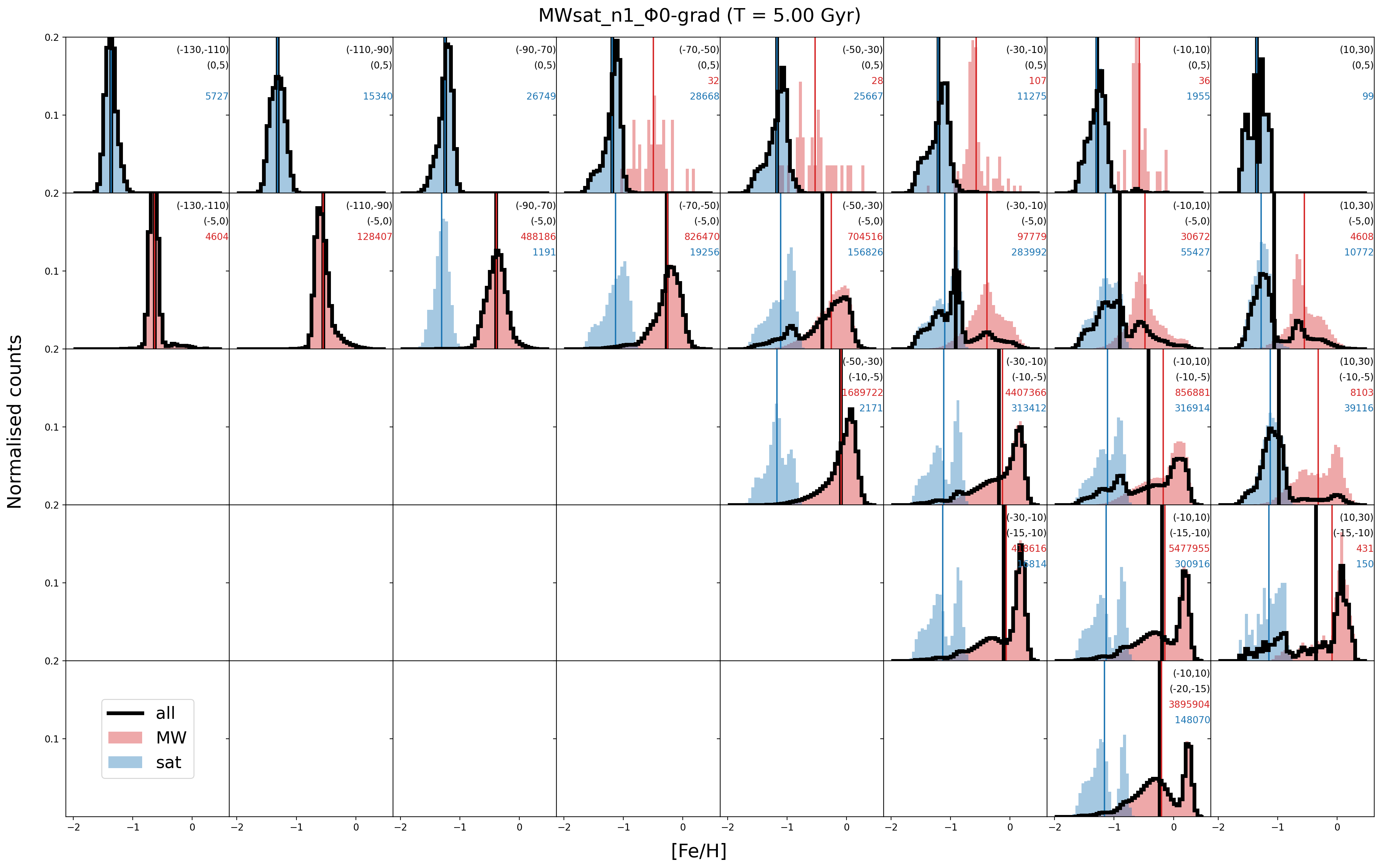}
    \caption{Same as Figure \ref{fig:elzmdf}, but for the case of $\phi_{orb}=0^{\circ}$. The upper panel shows the case of an initial vertical metallicity gradient only and the lower panel the one with an additional initial radial metallicity gradient.}
    \label{fig:elzmdf0}
\end{figure*}
\begin{figure*}%[H]
    \includegraphics[width=1\linewidth]{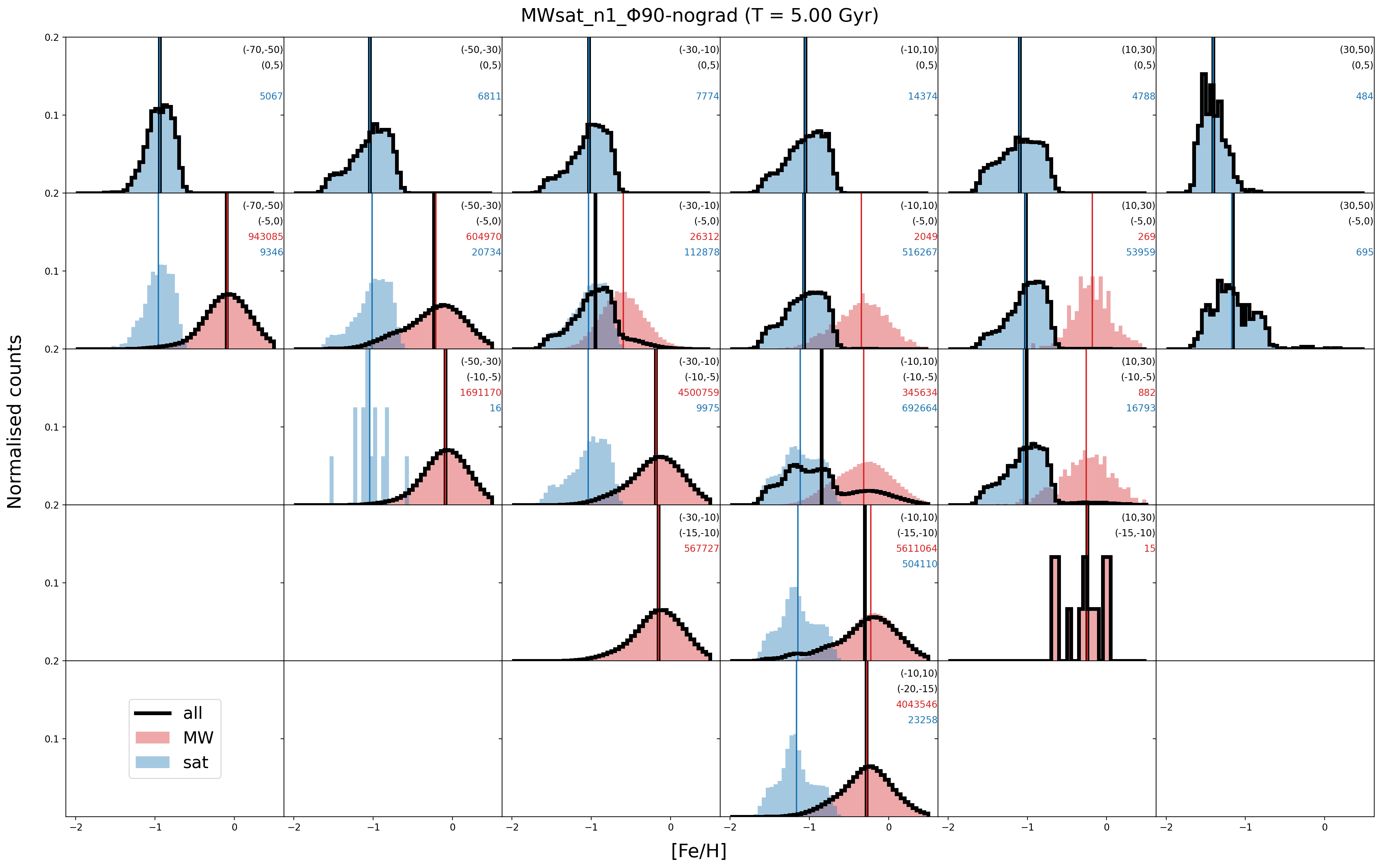}    
    \includegraphics[width=1\linewidth]{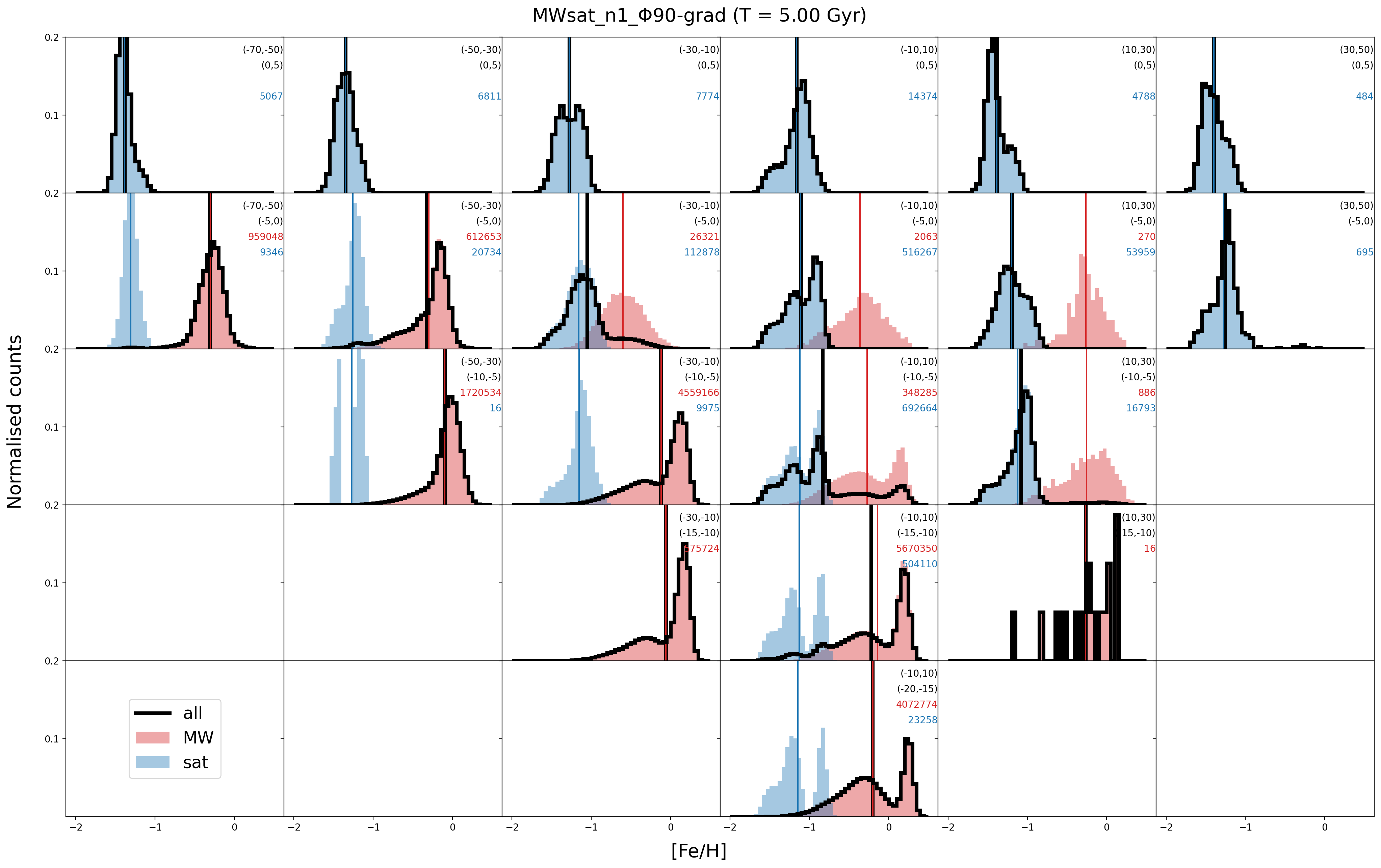}
    \caption{Same as Figure \ref{fig:elzmdf}, but for the case of $\phi_{orb}=90^{\circ}$. The upper panel shows the case of an initial vertical metallicity gradient only and the lower panel the one with an additional initial radial metallicity gradient.}
    \label{fig:elzmdf90}
\end{figure*}
\begin{figure*}%[H]
    \includegraphics[width=1\linewidth]{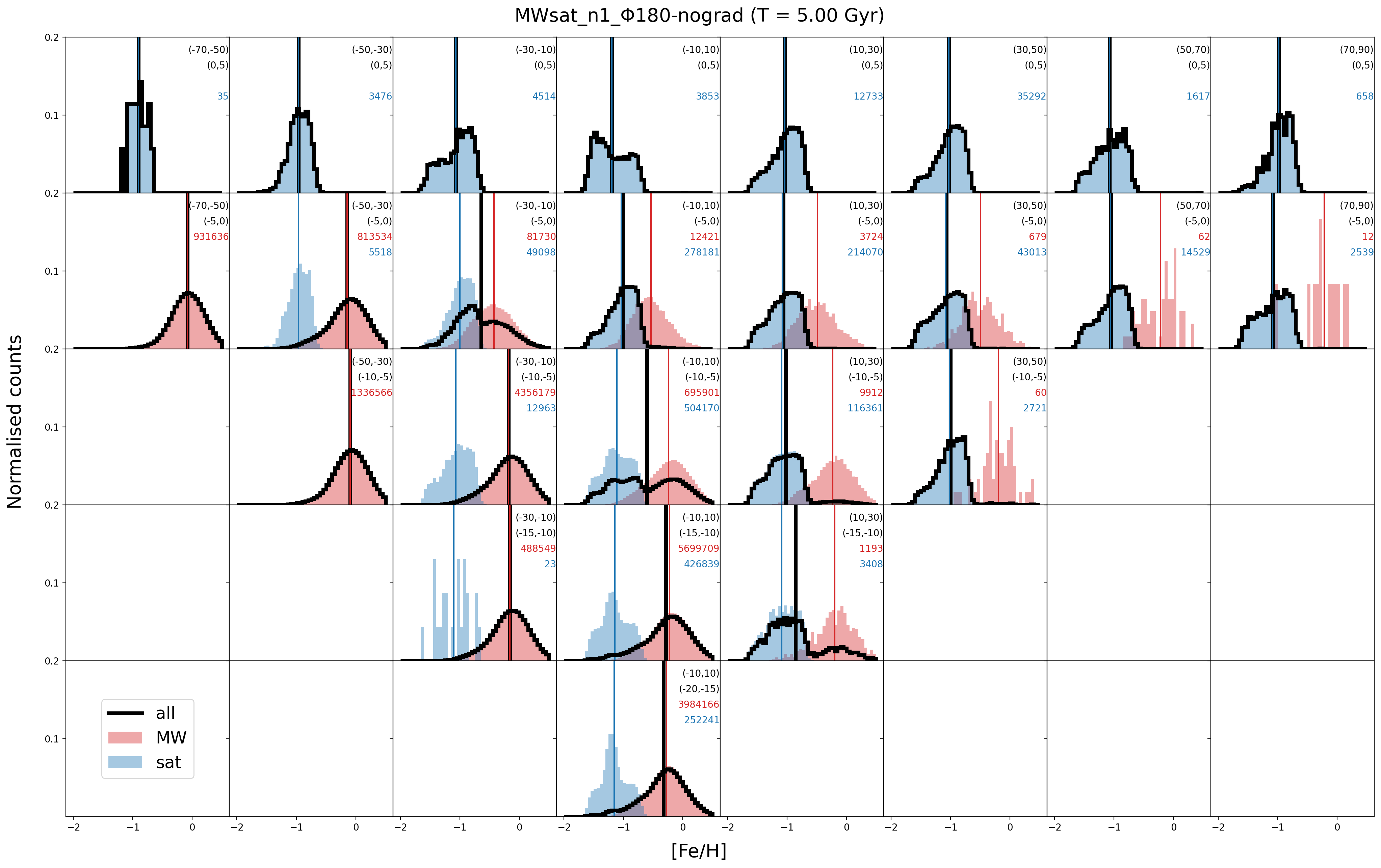}    
    \includegraphics[width=1\linewidth]{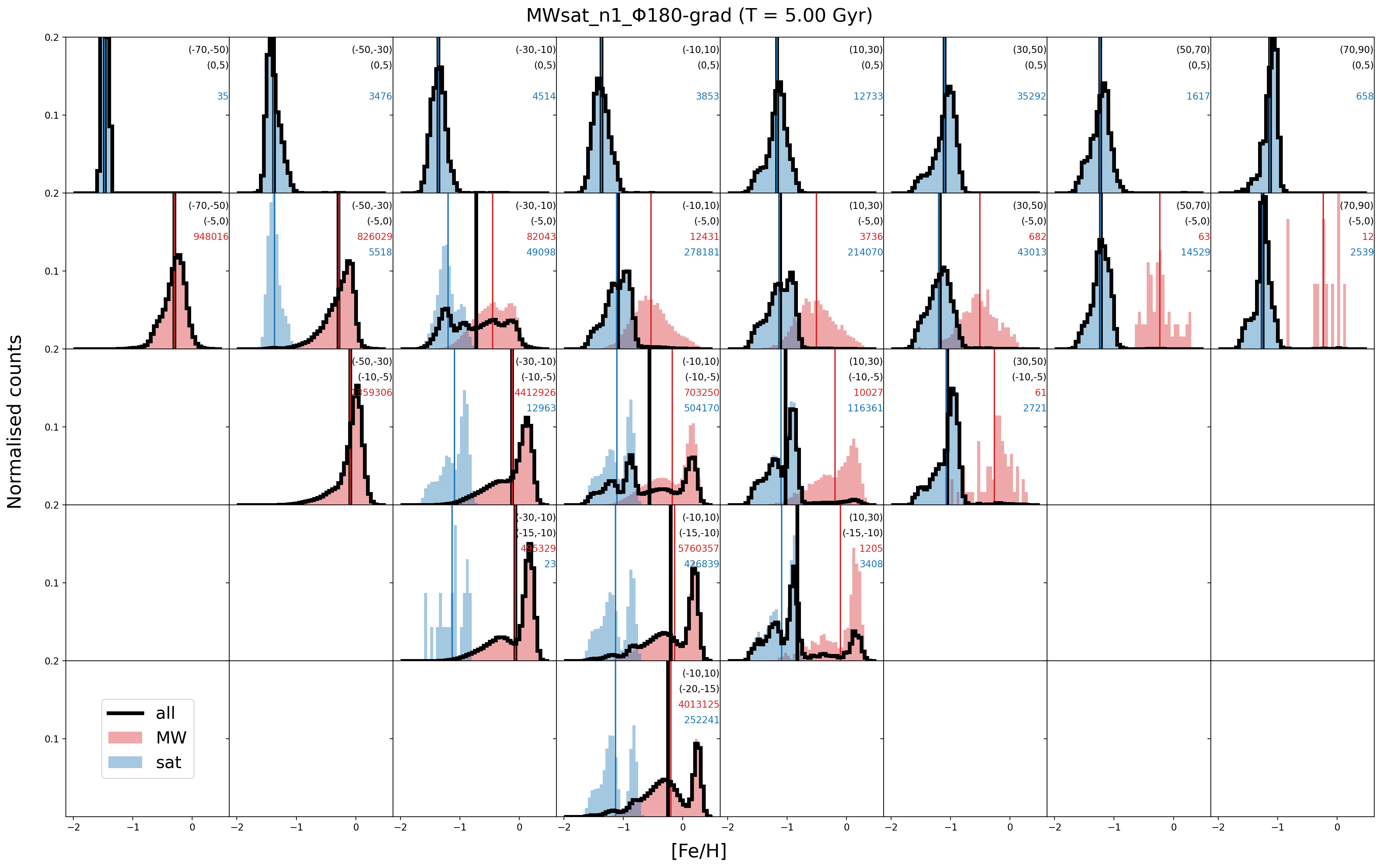}
    \caption{Same as Figure \ref{fig:elzmdf}, but for the case of $\phi_{orb}=180^{\circ}$. The upper panel shows the case of an initial vertical metallicity gradient only and the lower panel the one with an additional initial radial metallicity gradient.}
    \label{fig:elzmdf180}
\end{figure*}

\section{Fitting the metallicity gradient in the $E-L_z$ space as a function of energy}
\label{sec:fit}

\begin{figure*}%[H]
    \includegraphics[width=1\linewidth]{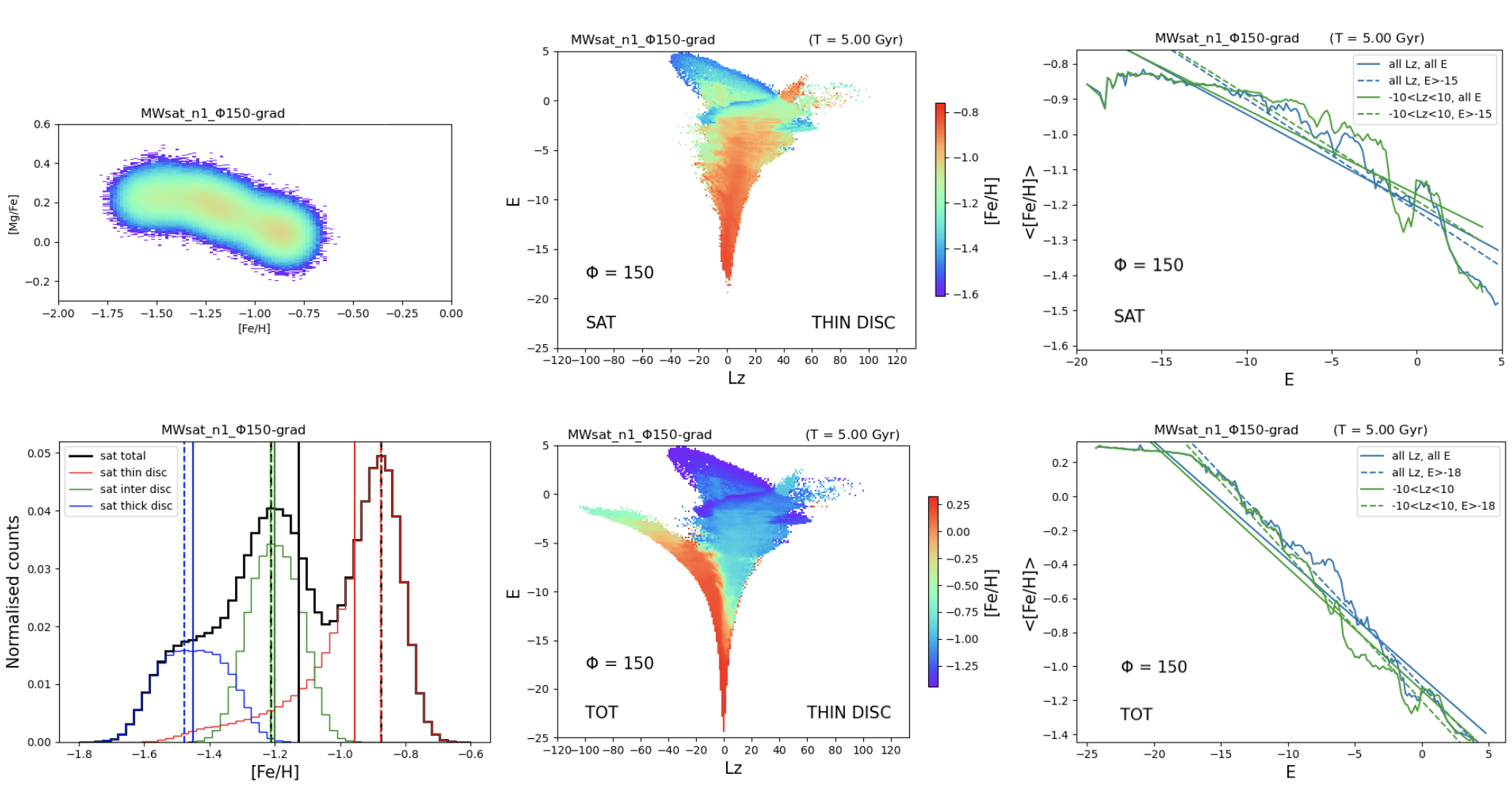} 
    \caption{Description of the method for fitting the metallicity gradient in the $E-L_z$ space as a function of energy. The first column shows the distribution in the [Mg/Fe]-[Fe/H] space and the MDF for every satellite disc component, with their means (peaks) in solid (dashed) lines. The second column shows instead the final distributions in the $E-L_z$ space of the satellite (and in situ) thin disc stars in the top (bottom) panel, colour-coded at each point with the mean metallicity. Finally, the third column shows the mean metallicity (<[Fe/H]>) as a function of energy (E) of the satellite (and in situ) thin disc stars in the top (bottom) panel.}
    \label{fig:fit}
\end{figure*}

As far as the relation mentioned in Section \ref{sec:relgrad} between the metallicity gradient in the satellite and in the $E-L_z$ space is concerned, we proceeded in the following way. In Figure \ref{fig:fit}, we report an example of the analysis we performed for every value of the slope of the radial metallicity gradient in the satellite thin disc before the interaction ($f_{satthin} \in [-0.06, 0.02]$ dex/kpc), in the case of $f_{satthin}=0.05$ dex/kpc and $f_{MWthin}=0.02$ dex/kpc, which is also analysed in the rest of the paper. The first column shows the distribution in the [Mg/Fe]-[Fe/H] space and the MDF for every satellite disc component, with their means (peaks) in solid (dashed) lines. The second column shows instead the final distributions in the $E-L_z$ space of the satellite (and in situ) thin disc stars in the top (bottom) panel, in the case of the initial inclination of the satellite orbital plane of $\phi_{orb}=150^{\circ}$ also analysed in previous sections. 

We computed the mean metallicity (<[Fe/H]>) as a function of energy (E), reported in the third column of Figure \ref{fig:fit} for only satellite thin disc stars in the top panel and for satellite and in situ thin disc stars in the bottom one. We then performed a linear fit considering four different cases. The cases we considered differ in the first place in taking into account stars either with all values of angular momentum ($L_z$) or only in the interval $-10\le L_z \le 10$ ($L_z\sim0$), and secondly in taking into account stars either with all values of energy (E) or performing a cut at low energy values ($E\ge -15$ for satellite stars only, $E\ge -18$ for satellite stars together with in situ ones), in order to avoid a plateau of the bulk of very bound stars having almost the same iron abundance. From the fits, we derived the slopes (d<[Fe/H]>/dE) that we report in Tables \ref{tab:fitsat} and \ref{tab:fittot}. 

We then selected the case of all $L_z$ values, together with the cut in energy, in order to restrict the analysis to the monotonic behaviour, and related its slope to $f_{satthin}$ in Section \ref{sec:relgrad}.

\begin{table}%[h]
    \centering
    \caption{Slopes (d<[Fe/H]>/dE) of the linear fit of the mean metallicity of satellite thin disc stars as a function of energy, in the four cases described in Appendix \ref{sec:fit}.}
    \label{tab:fitsat}    
    \begin{tabular}{|c|c c|}
         \hline
         & all E & $E\ge -15$ \\\hline
         all $L_z$ & -0.026 & -0.032\\
         $-10\le L_z \le 10$ & -0.024 & -0.30\\
         \hline
    \end{tabular}
\end{table}

\begin{table}%[h]
    \centering
    \caption{Same as Table \ref{tab:fitsat}, but for the cases of satellite and in situ thin disc stars.}
    \label{tab:fittot}
    \begin{tabular}{|c|c c|}
         \hline
         & all E & $E\ge -18$ \\\hline
         all $L_z$ & -0.069 & -0.083\\
         $-10\le L_z \le 10$ & -0.072 & -0.086\\
         \hline
    \end{tabular}
\end{table}

%\begin{figure}%[H]
%    \includegraphics[width=1\linewidth]{paper_fig_no/A5_mdf.png}    
%    \caption{}
%    \label{fig:}
%\end{figure}

%\begin{figure*}%[H]
%    \includegraphics[width=1\linewidth]{paper_fig_no/A5_fit.png} 
%    \caption{}
%    \label{fig:}
%\end{figure*}

\end{appendix}

%\end{comment}

%--------------------------------------------------------------------
\end{document}